\documentclass[aps,prl,preprint,floatfix]{revtex4-1}
\usepackage{graphicx}
\usepackage{comment}
\usepackage{amsmath}
\usepackage[left=1.5cm,right=1.5cm,top=2.0cm,bottom=2.0cm]{geometry}
\usepackage{braket}
\usepackage{dingbat}

\begin{document}

\title{Resonant theory of kinetic ballooning modes in general toroidal geometry}

\author{P.~Mulholland$^1$}
\author{A.~Zocco$^2$}
\author{M.~C.~L.~Morren$^1$}
\author{K.~Aleynikova$^2$} 
\author{M.~J.~Pueschel$^{3,1,4}$} 
\author{J.~H.~E.~Proll$^{1,2}$} 
\author{P.~W.~Terry$^5$}

\affiliation{
$^1$Eindhoven University of Technology, 5600 MB Eindhoven, The Netherlands\\ 
$^2$Max-Planck-Institut f{\"u}r Plasmaphysik, 17491 Greifswald, Germany\\ 
$^3$Dutch Institute for Fundamental Energy Research, 5612 AJ Eindhoven, The Netherlands\\
$^4$Department of Physics \& Astronomy, Ruhr-Universit{\"a}t Bochum, 44780 Bochum, Germany\\
$^5$University of Wisconsin-Madison, Madison, WI 53706, USA\\
}

\begin{abstract}
	The linear theory of the kinetic-ballooning-mode (KBM) instability is extended to capture a weakly-driven regime in general toroidal geometry where the destabilization is caused by the magnetic-drift resonance of the ions. Such resonantly-destabilized KBMs are characterized by broad eigenfunctions along the magnetic field line and near-marginal positive growth rates, even well below the $\beta$ threshold of their non-resonant counterparts. This unconventional (or sub-threshold) KBM, when destabilized, has been shown to catalyze an enhancement of turbulent transport in the Wendelstein 7-X (W7-X) stellarator \cite{Mulholland23, Mulholland25}. Simplifying the energy dependence of key resonant quantities allows for an analytical treatment of this KBM using the physics-based ordering from the more general equations of Tang, Connor, and Hastie \cite{Tang80}. Results are then compared with high-fidelity gyrokinetic simulations for the (st)KBM in W7-X and the conventional KBM in a circular tokamak at both high and low magnetic shear, where good agreement is obtained in all cases. This reduced KBM model provides deeper insight into (sub-threshold) KBMs and their relationship with geometry, and shows promise for aiding in transport model development and geometry-based turbulence optimization efforts going forward. 
\end{abstract}

\maketitle

\newpage

\section{Introduction} \label{sect:intro}
	
Achieving high normalized electron pressure $\beta \equiv \beta_\mathrm{e} = 8\pi n_\mathrm{0e} T_\mathrm{0e}/B_\mathrm{ref}^2$ is paramount for the successful operation of a future magnetic-confinement-fusion (MCF) power plant. Here, $n_\mathrm{0e}$ and $T_\mathrm{0e}$ respectively are the background electron density and temperature, and $B_\mathrm{ref}$ is the reference magnetic field strength on axis. High $\beta$ provides the necessary conditions for efficent energy generation from fusion reactions, but such high-pressure plasmas are prone to hosting electromagnetic instabilities, such as kinetic ballooning modes (KBMs). These instabilities modify turbulent behavior and facilitate unwanted heat and particle transport, lowering the energy-confinement time of the fusion reactor. KBMs are expected to thrive in high-$\beta$ regimes near the ideal MHD (iMHD) threshold $\beta_\mathrm{crit}^\mathrm{MHD}$. Both KBMs and iMHD ballooning modes are pressure-gradient driven instabilities, but KBMs differ from their iMHD counterpart in that they are modified by kinetic effects such as magnetic-drift resonances, finite-Larmor-radius (FLR) stabilization, trapped particles, Landau resonances, and collisions. These kinetic effects can have both destabilizing and stabilizing influences on the instability \cite{Hastie81, Cheng82a, Cheng82b, Hirose94, Hirose95, Hirose96}. 

Many theoretical studies of KBMs in toroidal geometry followed from the development of the ballooning-mode transformation, first detailed in Ref.~\cite{Connor78, Connor79}, which constitutes a procedure for reconciling the long parallel wavelength characteristic of plasma instabilities with periodicity of the geometry in a sheared toroidal magnetic field. Utilizing this procedure, the theory of KBMs in general toroidal geometry was developed in the seminal works of Ref.~\cite{Antonsen80, Tang80}. Following this, the influence of kinetic effects was studied in greater detail, including the impact of wave-particle resonances on KBMs \cite{Hastie81, Cheng82a, Cheng82b, Hirose94, Hirose95, Hirose96}. Further theoretical studies of KBMs in tokamak geometry can be found in Ref.~\cite{Tang85, Kotschenreuther86, Hong89a, Hong89b, Biglari91, Kim93, Tsai93, Zonca96, Zonca99, Zonca06, Aleynikova17, Zocco18b}. To support and advance this research, extensive numerical studies of electromagnetic instabilities and turbulence have been carried out both in tokamaks \cite{Snyder99thesis, Pueschel08, Pueschel10a, Bierwage10a, Bierwage10b, Belli10, Bierwage10a, Bierwage10b, Ishizawa13a, Maeyama14, Aleynikova17, Aleynikova18, Kumar21, Chen23, Pueschel25} and stellarators \cite{Baumgaertel12, Ishizawa14a, Ishizawa15a, Ishizawa17, Aleynikova18, Ishizawa19b, McKinney21, Mishchenko21, Mishchenko22, Aleynikova22, Mishchenko23, Mulholland23, Mulholland25}. Recent studies in the Wendelstein 7-X (W7-X) stellarator have revealed the deleterious manifestation of unconventional KBMs far below $\beta_\mathrm{crit}^\mathrm{MHD}$ \cite{Mulholland23, Mulholland25}. This variant has been classified as the sub-threshold KBM (stKBM), characterized by a destabilization threshold $\beta_\mathrm{crit}^\mathrm{stKBM}$ well below $\beta_\mathrm{crit}^\mathrm{MHD}$, and a consistently low but finite growth rate until reaching the conventional KBM threshold $\beta_\mathrm{crit}^\mathrm{KBM}$, at which point the stKBM transitions to a conventional KBM rapidly destabilized by further increases in $\beta$. The stKBM may be subdominant to coexisting instabilities present at low $\beta$. Despite its low growth rate, this instability has been shown to facilitate an enhancement of turbulent transport at low $\beta \geq \beta_\mathrm{crit}^\mathrm{stKBM}$ in ion-temperature-gradient-driven (ITG) turbulence regimes \cite{Mulholland23, Mulholland25}. Consequently, controlling stKBMs is of urgent interest in order to successfully achieve high performance in future MCF devices. Before stKBMs can be adequately controlled, a deeper understanding of them must first be established. This work aims to expand the knowledge basis for stKBMs by developing KBM theory that accounts for magnetic-drift resonance effects in three-dimensional flux-tube geometry. A numerical implementation of this theory is then shown to be capable of quickly and reliably predicting the existence and behavior of both stKBMs and conventional KBMs in general toroidal geometry. 

In this work, gyrokinetic theory is developed to obtain a simplified KBM equation suitable for general toroidal geometry that retains sufficent physics to capture distinctions between conventional KBMs and iMHD-like ballooning modes, as well as stKBMs. Previous studies have shown that stKBM growth rates are slightly augmented by the absence of trapped particles and parallel magnetic-field fluctuations $\delta B_\parallel$, but these effects need not be present for stKBMs to be unstable \cite{Mulholland25}. These findings have informed the simplifications made in the theory developed here: kinetic effects such as finite-Larmor-radius (FLR) stabilization and wave-particle resonances from passing ions subject to the magnetic-curvature drift (i.e., ion magnetic-drift resonances or toroidal resonances) are retained, while effects due to trapped particles, Landau resonances, and compressional magnetic fluctuations are neglected. The absence of $\delta B_\parallel$ is partially compensated for by replacing the $\nabla B$-drift frequency $\omega_{\nabla B}$ with the magnetic-curvature-drift frequency $\omega_\kappa$ (see definitions in Table~\ref{table:table_2}) \cite{Zocco15, Aleynikova22}. A collisionless plasma is assumed. The resulting KBM equation is reformulated as a nonlinear eigenvalue problem (i.e., having nonlinear dependence on the eigenvalue), amenable to numerical evaluation to obtain self-consistent (st)KBM eigenvalues and eigenvectors in any flux-tube geometry. Applying this approach in W7-X reveals that ion magnetic-drift resonances are necessary to explain the low-$\beta$ destabilization of the stKBM. In geometries where stKBMs remain absent (such as a circular tokamak), introducing resonance effects can lead to a moderate reduction in the destabilization threshold of the conventional KBM (relative to its non-resonant counterpart), a softer onset (i.e., a less rapidly increasing growth rate with $\beta$ near marginality), and a reduced growth rate at higher $\beta$. In all cases, the resonant-destabilization threshold differs from the non-resonant-destabilization threshold only in the presence of a finite ion-temperature gradient ($\eta_\mathrm{i}\neq0$; see definition in Table~\ref{table:table_2}), and the separation between these thresholds is proportional to $\eta_\mathrm{i}$ (compare Refs.~\cite{Cheng82a, Cheng82b, Hirose94, Hirose95, Zonca96, Zonca99}). The results from this reduced KBM model are in good agreement with numerical studies using a high-fidelity gyrokinetic code, highlighting that the physics included here is sufficient to explain and predict the existence (or absence) of stKBMs. These studies indicate the emergence of stKBMs is attributed to a combination of geometric characteristics, including local/background magnetic shear and magnetic field-line curvature. This enhanced understanding of stKBMs marks the first step toward designing an stKBM-resilient stellarator, more suitable for high-performance scenarios.

The remainder of this article is structured as follows. In section~\ref{sect:simplified_kbm_theory}, an analytical theory of KBMs is outlined. In section~\ref{sect:kbm_eigenvalue_problem}, the KBM equation is reformulated as a nonlinear eigenvalue problem (having nonlinear dependence on the eigenvalue), providing a means to obtaining self-consistent eigenvalues and eigenvectors for (st)KBMs in general toroidal geometry. In section~\ref{sect:results}, results of the KBM model are presented and compared with high-fidelity simulations in the W7-X stellarator and a high/low-magnetic-shear tokamak, followed by a physical interpretation and discussion. In section~\ref{sect:conclusion}, the article is closed with a short summary of key results and an outlook.

\newpage

\section{KBM theory} \label{sect:simplified_kbm_theory}

In this section, an eigenmode equation for the KBM is developed in general toroidal geometry; this equation is a radially-local (i.e., at fixed radial position $r/a$, with minor-radial coordinate $r$ and minor radius $a$) second-order ODE for the eigenmode along the magnetic field line. This derivation starts from the KBM theory presented in Ref.~\cite{Tang80}, and leads to the general-toroidal-geometry equivalent of the KBM equation derived in Ref.~\cite{Cheng82a}, where the latter was formulated for circular-tokamak geometry. The resulting equation includes ion-magnetic-drift resonaces and FLR effects, but neglects trapped particles, Landau resonances, collisions, and compressional magnetic fluctuations. Despite these reductions, the resulting equation is found to contain sufficient physics to capture important (st)KBM behavior in both tokamaks and stellarators. Following this derivation, the equation is presented in a dimensionless form amenable to numerical implementation.  




\subsection{Simplified KBM equation in general toroidal geometry}

Given the highly anisotropic character of plasma instabilities and turbulent fluctuations in fusion devices -- where correlation lengths parallel to the magnetic field are typcially 2-3 orders of magnitude greater than their perpendicular counterparts \cite{goerler11} -- it is convenient to employ a field-aligned coordinate system \cite{Helander14}. A general coordinate system is first introduced here followed by a related system utilized in flux-tube simulations \cite{Xanthopoulos09}. To start, one can introduce $(\psi,\theta,\zeta)$, where $\psi$ is the toroidal magnetic flux within a magnetic surface and acts as a radial variable, $\theta$ is the straight field-line poloidal angle, and $\zeta$ is the straight field-line toroidal angle \cite{Boozer81report,Beer95thesis,Faber18}. A field-line-labelling coordinate (i.e., fixed for a given field line) can then be defined as $\alpha = q(\psi)\theta - \zeta$, where $q(\psi) = d\zeta/d\theta = \iota(\psi)^{-1}$ is the safety factor and $\iota(\psi)$ is the rotational transform. This allows one to write the magnetic field in Clebsch form, 
\begin{equation}
	\mathbf{B} = \nabla \psi \times \nabla \alpha
\end{equation}

or alternatively
\begin{align}
	\mathbf{B} &= \nabla \psi \times \nabla \theta - \iota \nabla \psi \times \nabla \zeta \\
		   &= \nabla \psi \times \nabla \theta + \nabla \zeta \times \nabla \chi\\
\end{align}

where $\chi$ is the poloidal flux such that $\iota(\psi) = d\chi/d\psi$. In this coordinate system, the Jacobian is given by
\begin{equation}
	J = (\nabla \psi \times \nabla \theta \cdot \nabla \zeta)^{-1} = (\mathbf{B} \cdot \nabla \zeta)^{-1}. 
\end{equation}

In order for the following discussion to be consistent with the numerical flux-tube approach used by \textsc{Gene} \cite{Jenko00}, the following field-aligned coordinate system $(x,y,z)$ is used \cite{Xanthopoulos09} 
\begin{align}
	x &= L_\mathrm{ref}\sqrt{s} \\
	y &= L_\mathrm{ref} \frac{\sqrt{s_0}}{q_0} \alpha \\
	z &= \theta	
\end{align}

where $s=\psi/\psi_\mathrm{edge}$ is the normalized toroidal flux indicating radial position, $s_0$ is the radial position of a chosen flux surface (center of the flux-tube domain), $L_\mathrm{ref}$ is the macroscopic reference length (typically chosen to be the minor radius $a$, the major radius $R$, or the connection length $L_\mathrm{c} = q_0 R$, where $q_0 = q(s_0)$ is the safety factor on a given flux surface), and $\theta$ is a field-line-following poloidal angle coordinate (analogous to the ballooning angle \cite{Beer95thesis,Faber18}). In this coordinate system, the normalized magnetic field and Jacobian are
\begin{align}
	\hat{\mathbf{B}} &\equiv \frac{\mathbf{B}}{B_\mathrm{ref}}= \nabla x \times \nabla y, \label{B_xyz}\\
	J &= (\nabla x \times \nabla y \cdot \nabla z)^{-1} = (\hat{\mathbf{B}} \cdot \nabla z)^{-1}. \label{J_xyz}
\end{align} 

With the appropriate coordinate system established, the simplified KBM equation can now be derived. For the theory presented here, focus is given to the intermediate-frequency regime outlined in Ref.~\cite{Tang80} 
\begin{equation}
\omega_\mathrm{b,i}, \omega_\mathrm{t,i} \ll \omega \ll \omega_\mathrm{b,e}, \omega_\mathrm{t,e} \ ,\label{eq:intermediate_frequency_regime}
\end{equation}

where $\omega = \omega_\mathrm{r} + i\gamma$ is the complex eigenfrequency of the KBM with real frequency $\omega_\mathrm{r}$ and growth rate $\gamma$, and $\omega_{\mathrm{b},j}$ and $\omega_{\mathrm{t},j}$ respectively are the bounce and transit frequencies of particle species $j$. The relevant expansion parameter is $\epsilon = [v_{\mathrm{th,i}}/(\omega L_\mathrm{ref})]^2 \ll 1$, where $v_{\mathrm{th,i}} = \sqrt{2 T_\mathrm{0i}/m_\mathrm{i}}$ is the ion thermal velocity with ion temperature $T_\mathrm{0i}$ and ion mass $m_\mathrm{i}$. Specific expressions required for the following derivation are presented in Table~\ref{table:table_1}, and further definitions are provided in Table~\ref{table:table_2}.  


\begin{table}
\centering
\begin{tabular}{ p{10cm} p{8cm} }
\hline
\hline
\centering $\mathbf{B} = B\mathbf{b}$ & \text{Magnetic flux density}\\ 
\centering $c$ & \text{Speed of light}\\
\centering $q_\mathrm{i} = - q_\mathrm{e} = e$ & \text{Ion/electron charge}\\
\centering $\omega_{\mathrm{c}j} = \frac{q_jB}{m_jc}$ & \text{Cyclotron frequency of particle species $j$}\\
\centering $\mathbf{k} = \mathbf{k}_\parallel + \mathbf{k}_\perp$ & \text{Eigenmode wavenumber}\\
\centering $\mathbf{r} = \mathbf{R}_j - \frac{\mathbf{v} \times \mathbf{b}}{\omega_{\mathrm{c}j}}$ & \text{Particle position (with guiding center $\mathbf{R}_j$)} \\ 
\centering $\phi$ & \text{Perturbed electrostatic potential} \\
\centering $A_\parallel$ & \text{Perturbed parallel magnetic-vector potential} \\
\centering $\delta B_\parallel$ & \text{Perturbed parallel magnetic field} \\
\centering $\chi = \phi - \mathbf{v} \cdot \frac{\mathbf{A}}{c}$ & \text{Gyrokinetic potential} \\
\centering $f_j(\mathbf{r},\mathbf{v},t) = F_{\mathrm{M}j}(\mathbf{v}) \left(1-\frac{q_j \phi(\mathbf{r},t)}{T_{0j}}\right) + h_j(\mathbf{R}_j,\mathbf{v},t)$ & \text{Full distribution function} \\
\centering $h_j(\mathbf{R}_j,\mathbf{v},t) = \hat{h}_j \mathrm{e}^{i(\mathbf{k} \cdot \mathbf{R}_j - \omega t)}$ & \text{Non-adiabatic distribution function} \\
\centering $\langle \phi(\mathbf{r},\mathbf{v},t) \rangle_{\mathbf{R}_j} = \hat{\phi} \mathrm{e}^{i(\mathbf{k \cdot \mathbf{R}_j} - \omega t)} J_0 \left(\frac{k_\perp v_\perp}{\omega_{\mathrm{c}j}} \right)$ & \text{Gyro-averaging at fixed $\mathbf{R}$} \\
\centering $\langle h_j(\mathbf{R}_j,\mathbf{v},t) \rangle_{\mathbf{r}} = \hat{h}_j \mathrm{e}^{i(\mathbf{k \cdot \mathbf{r}} - \omega t)} J_0 \left(\frac{k_\perp v_\perp}{\omega_{\mathrm{c}j}} \right) $ & \text{Gyro-averaging at fixed $\mathbf{r}$}\\
\centering $\int d^3 v \, h_j(\mathbf{R}_j,\mathbf{v},t) = \int d^3 v \, \langle h_j(\mathbf{R}_j,\mathbf{v},t) \rangle_{\mathbf{r}}$ & \text{Equivalent integrals}\\
\hline
\hline
\end{tabular}
\caption{Definitions of quantities in gyrokinetic KBM theory \cite{Howes06}.}
\label{table:table_1}
\end{table}

The linearized collisionless gyrokinetic equation \cite{Rutherford68} governing the evolution of the non-adiabatic part of the perturbed distribution function $h_j$ for particle species $j$ (i.e., the distribution of charged rings responding to the perturbed fields) is  
\begin{equation}
	v_\parallel \nabla_\parallel \hat{h}_j - i(\omega - \omega_{\mathrm{D}j}) \hat{h}_j = -i\frac{q_j F_{\mathrm{M}j}}{T_{0j}} (\omega - \omega_{*j}^T) \langle \hat{\chi} \rangle_{\mathbf{R}_j}. \label{eq:GKE} 
\end{equation}

where the gyro-averaged potential is
\begin{equation}
	\langle \chi \rangle_{\mathbf{R}_j} = \langle \hat{\chi} \rangle_{\mathbf{R}_j} \mathrm{e}^{i(\mathbf{k} \cdot \mathbf{R}_j - \omega t)} =  \left[ J_0 \left(\frac{k_\perp v_\perp}{\omega_{\mathrm{c}j}} \right) \left(\hat{\phi} - \frac{v_\parallel \hat{A}_\parallel}{c} \right) + \frac{J_1\left(\frac{k_\perp v_\perp}{\omega_{\mathrm{c}j}} \right)}{\left(\frac{k_\perp v_\perp}{\omega_{\mathrm{c}j}} \right)} \frac{m v_\perp^2}{q_j} \frac{\delta \hat{B}_\parallel}{B} \right] \mathrm{e}^{i(\mathbf{k} \cdot \mathbf{R}_j - \omega t)},
\end{equation}

the Maxwellian background distribution function takes the usual form
\begin{equation}
	F_{\mathrm{M}j} = n_0 \left(\frac{m_j}{2T_{0j}}\frac{1}{\pi} \right)^{3/2} e^{-E_{j}/T_{0j}} = n_0 \left(\frac{1}{v_{{\mathrm{th}j}}^2 \pi} \right)^{3/2} e^{-\hat{v}^2},
\end{equation}

where hat quantities (except for velocity terms) denote the slowly varying amplitude of the fluctuating quantity in question, and $J_0$ and $J_1$ respectively are the zeroth- and first-order Bessel functions. The remaining quantities are defined in Table~\ref{table:table_2}. Using $v_\parallel \nabla_\parallel h_j \to i k_\parallel v_\parallel h_j$, Eq.~(\ref{eq:GKE}) can be rearranged to obtain an expression for $\hat{h}_j$ 
\begin{equation}
	\hat{h}_j = \frac{q_j F_{\mathrm{M}j}}{T_{0j}} \left(\frac{\omega - \omega_{*j}^T}{\omega - \omega_{\mathrm{D}j} - k_\parallel v_\parallel}\right) \langle \hat{\chi} \rangle_{\mathbf{R}_j}, 
\end{equation}

\begin{table}
\centering
\begin{tabular}{ p{8cm} p{8.5cm} }
\hline
\hline
\centering $E_\mathrm{i} = \frac{1}{2} m_\mathrm{i} v^2$ & \text{Ion kinetic energy}\\
\centering $\hat{v}^2 \equiv \frac{v^2}{v_{\mathrm{th,i}}^2} = \frac{E_\mathrm{i}}{T_\mathrm{0i}}$ & \text{Normalized velocity}\\ 
\centering $L_g = -\left(\frac{d \ln(g_0)}{dx} \right)^{-1}$ & \text{Equilibrium-gradient scale length ($g \in n,T_j$)}\\
\centering $\eta_j = \frac{L_n}{L_{T_\mathrm{j}}}$ & \text{Gradient ratio}\\
\centering $\beta_\mathrm{ref} = \beta_\mathrm{e} \hat{B}^2$ & \text{Reference normalized plasma pressure}\\ 
\centering $c_\mathrm{s} = \frac{1}{\sqrt{2}}v_\mathrm{th,i} = \sqrt{\frac{T_\mathrm{0e}}{m_\mathrm{i}}}$ & \text{Ion sound speed}\\
\centering $\omega_{\mathrm{c,ref}} = \frac{q_\mathrm{i}B_\mathrm{ref}}{m_\mathrm{i}c}$ & \text{Reference ion-cyclotron frequency}\\
\centering $\rho_\mathrm{th,i} = \frac{v_{\mathrm{th,i}}}{\omega_{\mathrm{ci}}}$ & \text{Thermal ion gyro-radius}\\
\centering $\rho_\mathrm{ref} = \frac{v_{\mathrm{th,i}}}{\omega_{\mathrm{c,ref}}}$ & \text{Reference ion gyro-radius}\\
\centering $\rho_\mathrm{s} = \frac{c_{\mathrm{s}}}{\omega_{\mathrm{c,ref}}}$ & \text{Ion sound gyro-radius}\\
\centering $\tau = \frac{T_\mathrm{0e}}{T_\mathrm{0i}}$ & \text{Temperature ratio}\\
\centering $\frac{d\beta_\mathrm{tot}}{dx} \equiv \beta_\mathrm{tot}' = -\frac{1}{L_\mathrm{ref}} \sum \limits_j \beta_\mathrm{ref} \left(\frac{L_\mathrm{ref}}{L_{n_j}} + \frac{L_\mathrm{ref}}{L_{T_j}} \right)$ & \text{Total normalized pressure gradient}\\
\centering $g^{ij} = \nabla x^i \cdot \nabla x^j$ & \text{Components of contravariant metric tensor}\\
\centering $\mathcal{L}_y = \frac{L_\mathrm{ref}}{B_\mathrm{ref}}\left(\partial_x B - \left(\frac{g^{xy}g^{yz} - g^{yy}g^{xz}}{g^{xx}g^{yy} - g^{yx}g^{xy}}\right) \partial_z B\right) $ & \text{Binormal curvature for $\nabla B$ drift}\\
\centering $\mathcal{K}_y = \mathcal{L}_y + \frac{L_\mathrm{ref} \beta_\mathrm{tot}'}{2\hat{B}}$ & \text{Binormal curvature for curvature drift}\\
\centering $\omega_{\nabla B} = \frac{1}{2}k_y \rho_\mathrm{ref} \frac{v_\mathrm{th,i}}{L_\mathrm{ref}} \mathcal{L}_y$ & \text{$\nabla B$-drift frequency}\\
\centering $\omega_\kappa = \frac{1}{2}k_y \rho_\mathrm{ref} \frac{v_\mathrm{th,i}}{L_\mathrm{ref}} \mathcal{K}_y$ & \text{Curvature-drift frequency}\\
\centering $\omega_{*\mathrm{i}} = - \frac{1}{2} k_y \rho_\mathrm{ref} \frac{v_{\mathrm{th,i}}}{L_n} = - \frac{\omega_{*\mathrm{e}}}{\tau}$ & \text{Ion diamagnetic frequency}\\
\centering $\omega_{*\mathrm{i}}^T = \omega_{*\mathrm{i}}\left[ 1 + \eta_\mathrm{i} \left( \frac{E_\mathrm{i}}{T_\mathrm{0i}} - \frac{3}{2} \right)\right]$ & \text{Energy-dependent ion diamagnetic frequency}\\
\centering $\omega_\mathrm{Di} = 2\left(\omega_\kappa	\hat{v}_\parallel^2 + \omega_{\nabla B} \frac{\hat{v}_\perp^2}{2} \right)$ & \text{Magnetic-curvature drift frequency}\\
\hline
\hline
\end{tabular}
\caption{Definitions of quantities in Eq.~(\ref{eq:GKE}) and terms therein.}
\label{table:table_2}
\end{table}

The aforementioned simplifications are made from the outset, where the effects of $\delta B_\parallel$ and trapped particles are neglected. Given the intermediate-frequency regime (\ref{eq:intermediate_frequency_regime}), the mode frequency is assumed to be much larger than the ion transit frequency $|\omega| \gg k_\parallel v_\mathrm{th,i}$ and much smaller than the electron transit frequency $|\omega| \ll k_\parallel v_\mathrm{th,e}$. The parallel current from the ions is found to be ignorable, and is largely carried by the electrons \cite{Hirose95}. For the electrons, FLR effects can be ignored. The perturbed distribution functions for the ions and electrons respectively are then 

\begin{align}
	\delta f_\mathrm{i} &= - \frac{q_\mathrm{i}}{T_\mathrm{0i}} \left( F_\mathrm{Mi} \phi - \hat{h}_\mathrm{i} \mathrm{e}^{i(\mathbf{k} \cdot \mathbf{R}_\mathrm{i} - \omega t)} \right) \nonumber\\ 
			    &= - \frac{e}{T_\mathrm{0i}} \left( F_\mathrm{Mi} \phi - F_\mathrm{Mi} \left(\frac{\omega - \omega_\mathrm{*i}^T}{\omega - \omega_{\mathrm{Di}}}\right) J_0 \left(\frac{k_\perp v_\perp}{\omega_{\mathrm{ci}}} \right) \left(\hat{\phi} - \frac{v_\parallel \hat{A}_\parallel}{c} \right) \mathrm{e}^{i(\mathbf{k} \cdot \mathbf{R}_\mathrm{i} - \omega t)} \right), \\
	\delta f_\mathrm{e} &= - \frac{q_\mathrm{e}}{T_\mathrm{0e}} \left( F_\mathrm{Me} \phi - \hat{h}_\mathrm{e} \mathrm{e}^{i(\mathbf{k} \cdot \mathbf{R}_\mathrm{e} - \omega t)} \right) \nonumber\\ 
			    &= \frac{e}{T_\mathrm{0e}} \left( F_\mathrm{Me} \phi - F_\mathrm{Me} \left(\frac{\omega - \omega_\mathrm{*e}^T}{- k_\parallel v_\parallel}\right) \left(\hat{\phi} - \frac{v_\parallel \hat{A}_\parallel}{c} \right) \mathrm{e}^{i(\mathbf{k} \cdot \mathbf{R}_\mathrm{e} - \omega t)} \right).
\end{align}

The perturbed densities for ions and electrons are obtained from $\delta n_j = \int d^3 v \, \delta f_j$ (where integration is performed at a fixed $\mathbf{r}$, where the charges reside). Using 
\begin{equation}
	\int d^3 v \, h_j = \int d^3 v \, \langle h_j \rangle_\mathbf{r} = \int d^3 v \, \hat{h}_j \mathrm{e}^{i(\mathbf{k \cdot \mathbf{r}} - \omega t)} J_0 \left(\frac{k_\perp v_\perp}{\omega_{\mathrm{ci}}} \right)
\end{equation}
leads to the perturbed ion density
\begin{align}
	\delta n_\mathrm{i} &= - \frac{e}{T_\mathrm{0i}} \left( \int d^3 v \, F_\mathrm{Mi} \phi - \int d^3 v \, h_\mathrm{i} \right)\nonumber \\ 
			    &= - \frac{e}{T_\mathrm{0i}} \left( n_0 \phi - \hat{\phi} \mathrm{e}^{i(\mathbf{k \cdot \mathbf{r}} - \omega t)} \int d^3 v \,F_\mathrm{Mi} J_0^2 \left(\frac{k_\perp v_\perp}{\omega_{\mathrm{ci}}} \right) \left(\frac{\omega - \omega_\mathrm{*i}^T}{\omega - \omega_{\mathrm{Di}}}\right) \right)\nonumber \\ 
			    &= - \frac{e n_0}{T_\mathrm{0i}} (1-Q) \phi 
\end{align}

where 
\begin{equation}
	Q \equiv \int d^3v \frac{F_{\mathrm{Mi}}}{n_0} J_0^2(\hat{v}_\perp \sqrt{2b}) \left(\frac{\omega - \omega_\mathrm{*i}^T}{\omega - \omega_\mathrm{Di}} \right) \label{eq:Q_unapproximated}
\end{equation}

with $b=(k_\perp \rho_\mathrm{th,i})^2/2$, and terms that are odd with respect to $v_\parallel$ vanish on integration. 

Similarly, for the electrons
\begin{align}
	\delta n_\mathrm{e} &= \frac{e}{T_\mathrm{0e}} \left( \int d^3 v \, F_\mathrm{Me} \phi - \int d^3 v \, h_\mathrm{e} \right) \nonumber \\
			    &= \frac{e}{T_\mathrm{0e}} \left( n_0 \phi - \left(\frac{\omega \hat{A}_\parallel}{k_\parallel c} \right)\mathrm{e}^{i(\mathbf{k \cdot \mathbf{r}} - \omega t)} \int d^3 v \, F_\mathrm{Me} \left(1 - \frac{\omega_\mathrm{*e}^T}{\omega}\right)  \right) \nonumber \\
			    &\simeq \frac{e n_0}{T_\mathrm{0e}} \left( \phi - \left(1 - \frac{\omega_\mathrm{*e}}{\omega}\right) \left(\frac{\omega A_\parallel}{k_\parallel c} \right) \right).
\end{align}

Utilizing the ballooning transform \cite{Connor78} in a similar fashion as Ref.~\cite{Hirose95}, one can relate 
\begin{equation}
	\nabla_\parallel = i k_\parallel = \frac{1}{J\hat{B}} \frac{\partial}{\partial \theta}
\end{equation}

such that one can define \cite{Tang80} 
\begin{equation}
	\psi_\parallel \equiv \frac{\omega A_\parallel}{k_\parallel c} \implies \frac{1}{J\hat{B}} \frac{\partial}{\partial \theta} \psi_\parallel = \frac{i\omega}{c}A_\parallel. 
\end{equation}

where $\psi_\parallel$ is the opposite-parity representation of $A_\parallel$. 

With the perturbed densities of both particle species now obtained, they can be inserted into the quasi-neutrality condition 
\begin{equation}
	\sum\limits_j q_j \delta n_j = e \left[\delta n_\mathrm{i} - \delta n_\mathrm{e} \right] = 0,
\end{equation}

giving
\begin{equation}
	\left( 1 - \frac{\omega_\mathrm{*e}}{\omega} \right) \hat{\psi}_\parallel = (1 + \tau - \tau Q)\hat{\phi}, \label{eq:quasi_neutrality_reduced}
\end{equation}

where the potentials have been expanded as plane waves, leaving only the amplitudes (hat terms).

The parallel component of Amp{\`e}re's law involves taking the parallel-velocity moment of $h_j$ 
\begin{equation}
	- \nabla_\perp^2 A_\parallel = \frac{4\pi}{c} \delta j_\parallel = \sum \limits_j \frac{4\pi}{c} q_j \int d^3 v \, v_\parallel \langle h_j \rangle_{\mathbf{r}}
\end{equation}

where the adiabatic part vanishes due to odd integration with respect to  $v_\parallel$. By considering only contributions from the electron current, this law reads \cite{Tang80} 
\begin{equation}
	\frac{L_\mathrm{ref}^2}{J\hat{B}} \frac{\partial}{\partial \theta} \left( \frac{b}{J\hat{B}} \frac{\partial}{\partial \theta} \hat{\psi}_\parallel \right) = \left(\frac{\omega}{\omega_\mathrm{A}}\right)^2 \left\{ \left(Q - 1 - \frac{\omega_\mathrm{*e}}{\omega \tau} \right) \hat{\phi} - \left[ 1 - \frac{\omega_\mathrm{*e}}{\omega} (1+\eta_\mathrm{e}) \right] \left( \frac{\omega_\kappa + \omega_\mathrm{B}}{\omega} \right) \hat{\psi}_\parallel \right\} \label{eq:parallel_current_reduced}
\end{equation}

where $\omega_\mathrm{A}^2 = v_{\mathrm{th,i}}^2/(L_\mathrm{ref}^2 \beta_\mathrm{i})$ is the Alfv{\'e}n frequency, and $\beta_\mathrm{i} = \beta_\mathrm{e}/\tau$ for equal particle-species density $n_\mathrm{0i} = n_\mathrm{0e}$. 

Combining Eqs.~(\ref{eq:quasi_neutrality_reduced}) and (\ref{eq:parallel_current_reduced}) yields 
\begin{align}
	\frac{L_\mathrm{ref}^2}{J\hat{B}} \frac{\partial}{\partial \theta} \left( \frac{b}{J\hat{B}} \frac{\partial}{\partial \theta} \hat{\psi}_\parallel \right) &= \left(\frac{\omega}{\omega_\mathrm{A}}\right)^2 \biggl\{ \left(Q - 1 - \frac{\omega_\mathrm{*e}}{\omega \tau} \right)(1 + \tau - \tau Q)^{-1}\left( 1 - \frac{\omega_\mathrm{*e}}{\omega} \right) \nonumber \\  
																		       & - \left[ 1 - \frac{\omega_\mathrm{*e}}{\omega} (1+\eta_\mathrm{e}) \right] \left( \frac{\omega_\kappa + \omega_\mathrm{B}}{\omega} \right) \biggr\}  \hat{\psi}_\parallel \nonumber \\ 
																		       &= \left(\frac{\omega}{\omega_\mathrm{A}}\right)^2 \left\{ \frac{Q - \alpha_\mathrm{0,i}}{1 + \tau - \tau Q} \alpha_\mathrm{0,e} - \alpha_\mathrm{1,e} \left( \frac{\omega_\kappa + \omega_\mathrm{B}}{\omega} \right) \right\}\hat{\psi}_\parallel , 
\end{align}

where the condensed notation 
\begin{equation}
\alpha_{m,j} = 1 - \frac{\omega_{*j}}{\omega}(1 + m \eta_j)
\end{equation}

has been used. 

The KBM eigenfunction $\hat{\psi}_\parallel$ can be rewritten in terms of a reduced scalar $\Phi$ \cite{Hirose95} in the following manner. Rearranging Eq.~\ref{eq:quasi_neutrality_reduced} gives 
\begin{equation}
	\hat{\psi}_\parallel = \frac{(1 + \tau - \tau Q)}{\left( 1 - \frac{\omega_\mathrm{*e}}{\omega} \right)} \hat{\phi} = \frac{(1 + \tau - \tau Q)}{\alpha_\mathrm{0,e}} \hat{\phi} \equiv \Phi.
\end{equation}

By making use of an approximative form \cite{Aleynikova17} of the ion integral $Q$ (for $\omega_\mathrm{Di}/|\omega| \sim \epsilon \ll 1$), 
\begin{align}
	Q &\approx 1 - \frac{\omega_\mathrm{*i}}{\omega} + \left(\frac{\omega_{\nabla B} + \omega_\kappa}{\omega} - b \right) \left[1 - \frac{\omega_\mathrm{*i}}{\omega}(1+\eta_i)\right] + \mathcal{O}(\epsilon^2) \label{eq:Q_fluid} \nonumber \\
	  &\approx \alpha_\mathrm{0,i} + \left(\frac{\omega_{\nabla B} + \omega_\kappa}{\omega} - b \right) \alpha_\mathrm{1,i} \nonumber \\
	  &\approx \alpha_\mathrm{0,i} + \mathcal{O}(\epsilon)
\end{align}

one obtains to zeroth order,
\begin{align}
	\Phi &= \frac{(1 + \tau - \tau Q)}{\alpha_\mathrm{0,e}} \hat{\phi} \approx \frac{(1 + \tau - \tau \alpha_\mathrm{0,i})}{\alpha_\mathrm{0,e}}\hat{\phi} = \hat{\phi}
\end{align}

such that $\Phi$ is equivalent to $\hat{\phi}$ with corrections of order $\epsilon \ll 1$.	  

The KBM equation is now rewritten in a dimensionless form that is suitable for numerical implementation, where changes in notation are detailed in Table~\ref{table:table_3}. Additional simplifications are made here: only eigenmodes centered at $k_x=0$ are considered, and due to the absence of parallel magnetic-field fluctuations, a common approximation \cite{Zocco15, Aleynikova18} of setting $\omega_{\nabla B} \to \omega_\kappa$ is made, which can act to partially compensate for the lack of destabilization normally induced by $\delta B_\parallel$ \cite{Zocco15, Aleynikova18}. 

\begin{table}
\centering
\begin{tabular}{ p{6cm} p{9cm} }
\hline
\hline
\centering $b = \frac{k_\perp^2 \rho_\mathrm{th,i}^2}{2} \approx \frac{1}{2} k_y^2 \rho_\mathrm{ref}^2 \frac{g^{yy}}{\hat{B}^2}$ & \text{FLR term (considering modes centered at $k_x=0$)}\\
\centering $\hat{J} = \frac{J}{L_\mathrm{ref}}$ & \text{Normalized Jacobian}\\
\centering $\epsilon_n = \frac{L_n}{L_\mathrm{ref}}$ & \text{Normalized inverse density gradient}\\ 
\centering $\Omega = \frac{\omega}{\omega_\mathrm{*e}}$ & \text{Mode frequency}\\
\centering $\Omega_\kappa = \frac{\omega_\kappa}{\omega_\mathrm{*e}} = \frac{\epsilon_n \mathcal{K}_y}{\tau}$ & \text{Curvature-drift frequency}\\ 
\hline
\hline
\end{tabular}
\caption{Definitions and normalizations for the dimensionless KBM equation (\ref{eq:dimensionless_kbm_eqn_1}).}
\label{table:table_3}
\end{table}


Applying these changes, the KBM equation becomes 
\begin{equation}
	\frac{1}{\hat{J} \hat{B}} \frac{\partial}{\partial \theta} \left( \frac{g^{yy}}{\hat{J}\hat{B}^3} \frac{\partial}{\partial \theta} \Phi \right) = \Omega^2 \frac{\beta_\mathrm{i}\tau^2}{2\epsilon_n^2} \left\{ \frac{Q - \alpha_\mathrm{0,i}}{1 + \tau - \tau Q} \alpha_\mathrm{0,e} - \alpha_\mathrm{1,e} \frac{2\Omega_\kappa}{\Omega}\right\} \Phi. \label{eq:dimensionless_kbm_eqn_1}
\end{equation}

This KBM equation differs from those presented in Refs.~\cite{Tang80,Hirose94} in several respects. Unlike the result of Ref.~\cite{Tang80}, Eq.~(\ref{eq:dimensionless_kbm_eqn_1}) omits the effect of parallel-magnetic-field fluctuations (and uses the aforementioned approximation $\omega_{\nabla B} \to \omega_\kappa$), and the influence of trapped particles is neglected. The result of Ref.~\cite{Hirose94} includes similar physics as Eq.~(\ref{eq:dimensionless_kbm_eqn_1}), but is only applicable to circular tokamak geometry; Eq.~(\ref{eq:dimensionless_kbm_eqn_1}) is suitable for general toroidal geometry, i.e., shaped tokamaks and stellarators. The physics contained in Eq.~(\ref{eq:dimensionless_kbm_eqn_1}) can be understood as follows. The left-hand side captures stabilizing effects such as shear stabilization via $\partial g^{yy}(\theta)/\partial \theta$ ($g^{yy}$ also appears as an argument of the Bessel functions in $Q$, where it contributes to FLR damping) and magnetic-field-line-bending stabilization manifest via the degree of eigenmode localization along the field line, measured by $|\partial^2 \Phi/\partial \theta^2|$. When this equation is solved, these stabilizing effects are balanced by the drives on the right-hand side: the normalized plasma pressure $\beta_\mathrm{i}$; normalized pressure gradients are contained in $\epsilon_n^{-1}$, $Q$, and $\alpha_{m,j}$; the geometric drive of bad curvature is contained in $\Omega_\kappa$ and is also manifest in $Q$; and lastly, wave-particle resonances associated with passing ions subject to the magnetic-curvature drift is provided by $Q$. It is worth mentioning that the right-hand side of this equation is not purely destabilizing; for instance, kinetic resonances can act to dampen KBM growth rates away from marginality (shown in section~\ref{sect:results}) and $Q$ includes FLR-stabilization terms, see Eqs.~(\ref{eq:Q_fluid}) and (\ref{eq:Q_resonant}). 

This equation can be evaluated using either a resonant framework where the ion integral $Q$ is treated resonantly, thus retaining ion-magnetic-drift resonances (i.e., toroidal resonances), or by using a non-resonant framework where $Q$ is evaluated assuming $\omega_\mathrm{Di}/|\omega| \ll 1$, shown in Eq.~(\ref{eq:Q_fluid}). The resonant treatment is preferable if one wishes to accurately resolve the KBM's behavior near marginality, whilst a non-resonant approximation is suitable when the KBM is strongly unstable \cite{Aleynikova17}. The non-resonant KBM equation reduces to an iMHD ballooning mode equation in the limit $k_y \to 0$. For the resonant treatment, a simplified form of the resonant ion integral is derived here, in agreement with the result reported in Ref.~\cite{Cheng82a}.  

Utilizing notation from Ref.~\cite{Cheng82a}, and in keeping with their convention of normalizing frequencies by the electron diamagnetic frequency $\omega_\mathrm{*e}$,
\begin{align*}
	F_0 &= F_\mathrm{Mi}\frac{v_\mathrm{th,i}^3}{n_0} = \frac{1}{\pi^{3/2}}e^{-\hat{v}^2}\\
	\Omega &\equiv \frac{\omega}{\omega_\mathrm{*e}}\\
	\Omega_\mathrm{*i}^T &\equiv \frac{\omega_\mathrm{*i}^T}{\omega_\mathrm{*e}} = \frac{\omega_\mathrm{*i}}{\omega_\mathrm{*e}} [1 + \eta_\mathrm{i}(\hat{v}^2 - 3/2)] = -\frac{1}{\tau}[1 + \eta_\mathrm{i}(\hat{v}^2 - 3/2)]\\
	\Omega_\mathrm{si} &\equiv \tau(\Omega - \Omega_\mathrm{*i}^T) = \tau \Omega + [1 + \eta_\mathrm{i}(\hat{v}^2 - 3/2)]\\
	\Omega_\mathrm{Di} &\equiv \frac{\omega_\mathrm{Di}}{\omega_\mathrm{*e}},
\end{align*}

such that $Q$ can be obtained from
\begin{align}
	\tau Q &= \tau \int d^3v \frac{F_\mathrm{Mi}}{n_0} J_0^2 (\sqrt{2b} \hat{v}_\perp) \frac{\omega_\mathrm{*e}}{\omega_\mathrm{*e}} \left(\frac{\omega - \omega_\mathrm{*i}^T}{\omega - \omega_\mathrm{Di}} \right) \nonumber \\ 
	       &= \int d^3\hat{v} F_0 J_0^2 (\sqrt{2b} \hat{v}_\perp) \left(\frac{\Omega_\mathrm{si}}{\Omega - \Omega_\mathrm{Di}} \right),
\end{align}

which aside from a different sign convention for $\Omega_\mathrm{Di}$ is equivalent to Eq.~(18) in Ref.~\cite{Cheng82a}.

To make this integral solvable analytically, the approximation is made of setting $\hat{v}_\perp^2 = 2 \hat{v}_\parallel^2$ in the denominator, which corresponds to a common curvature model in ITG theory \cite{Terry82}; this simplification is made for KBM theory in Ref.~\cite{Cheng82a}, which removes the resonance present in the $\hat{v}_\perp$ integral but retains it in the $\hat{v}_\parallel$ integral. As explained in Ref.~\cite{Cheng82a}, this approximation deforms the resonant surface in velocity space into a plane, which provides analytical tractability without significantly affecting the underlying magnetic-curvature-drift-resonance effects. 

The velocity-space integrals are expanded as
\begin{equation}
	\int d^3 \hat{v} = 2\pi \int\limits_{-\infty}^{+\infty} d \hat{v}_\parallel \int\limits_{0}^{+\infty} d \hat{v}_\perp \hat{v}_\perp, \\
\end{equation}

and the following relations will be utilized
\begin{align}
	\frac{1}{\sqrt{\pi}} \int\limits_{-\infty}^{+\infty} dx \frac{e^{-x^2}}{x-\zeta} &\equiv Z(\zeta) \label{eq:plasma_disp_function}\\
	\frac{1}{i \pi} \int\limits_{-\infty}^{+\infty} dx \frac{e^{-x^2}}{x-\zeta} &= \frac{1}{i\sqrt{\pi}} Z(\zeta) \equiv W(\zeta) \label{eq:faddeeva_function_1}\\
	\frac{2 \zeta}{i \pi} \int\limits_{0}^{+\infty} dx \frac{e^{-x^2}}{x^2-\zeta^2} &= \frac{\zeta}{i \pi} \int\limits_{-\infty}^{+\infty} dx \frac{e^{-x^2}}{x^2-\zeta^2} = W(\zeta) \label{eq:faddeeva_function_2} \\
	\int\limits_{0}^{+\infty} d \hat{v}_\perp \hat{v}_\perp e^{-\hat{v}_\perp^2} J_0^2 (\sqrt{2b} \hat{v}_\perp) &= \frac{1}{2} \Gamma_0(b) \\
	\int\limits_{0}^{+\infty} d \hat{v}_\perp \hat{v}_\perp^3 e^{-\hat{v}_\perp^2} J_0^2 (\sqrt{2b} \hat{v}_\perp) &= \frac{1}{2} [\Gamma_0(b) + b(\Gamma_1(b)-\Gamma_0(b))]
\end{align}

where Eq.~(\ref{eq:plasma_disp_function}) is the plasma dispersion function, Eqs.~(\ref{eq:faddeeva_function_1}) and (\ref{eq:faddeeva_function_2}) define the Faddeeva function, and $\Gamma_n(b) \equiv I_n(b) e^{-b}$, where $I_n(b)$ is the modified Bessel function of the first kind. Note that the integral representations for $Z(\zeta)$ and $W(\zeta)$ are evaluated only for $\textrm{Im}(\zeta)>0$. 

Using these definitions, one may write
\begin{align}
	       \tau Q &= \frac{2}{\sqrt{\pi}} \int\limits_{-\infty}^{+\infty} d \hat{v}_\parallel \left\{\frac{e^{-\hat{v}_\parallel^2}}{\Omega - 4 \Omega_\kappa \hat{v}_\parallel^2} \right\} \int\limits_{0}^{+\infty} d \hat{v}_\perp \hat{v}_\perp \left\{ e^{-\hat{v}_\perp^2} J_0^2 (\sqrt{2b} \hat{v}_\perp) \left([\tau \Omega + 1 + \eta_\mathrm{i}\hat{v}_\parallel^2 - \eta_\mathrm{i}3/2] + \eta_\mathrm{i}\hat{v}_\perp^2 \right) \right\}. 
\end{align}

Focusing only on the $\hat{v}_\perp$ integral
\begin{align}
	&\int\limits_{0}^{+\infty} d \hat{v}_\perp \hat{v}_\perp \left\{ e^{-\hat{v}_\perp^2} J_0^2 (\sqrt{2b} \hat{v}_\perp) \left([\tau \Omega + 1 + \eta_\mathrm{i}\hat{v}_\parallel^2 - \eta_\mathrm{i}3/2] + \eta_\mathrm{i}\hat{v}_\perp^2 \right) \right\} \nonumber \\ 
	   &{} = \frac{1}{2}\left(\eta_\mathrm{i}\hat{v}_\parallel^2 \Gamma_0(b) + [\tau \Omega + 1 - \eta_\mathrm{i}/2] \Gamma_0(b) + \eta_\mathrm{i} b(\Gamma_1(b)-\Gamma_0(b))\right). 
\end{align}

Reinserting this result into the full expression, one arrives at
\begin{align}
	 \tau Q &= \frac{2}{\sqrt{\pi}} \int\limits_{-\infty}^{+\infty} d \hat{v}_\parallel \left\{\frac{e^{-\hat{v}_\parallel^2}}{\Omega - 4 \Omega_\kappa \hat{v}_\parallel^2} \right\} \frac{1}{2}\left(\eta_\mathrm{i}\hat{v}_\parallel^2 \Gamma_0(b) + [\tau \Omega + 1 - \eta_\mathrm{i}/2] \Gamma_0(b) + \eta_\mathrm{i} b(\Gamma_1(b)-\Gamma_0(b))\right) \nonumber \\
		 &= \frac{1}{\sqrt{\pi}} \eta_\mathrm{i} \Gamma_0(b) \int\limits_{-\infty}^{+\infty} d \hat{v}_\parallel \hat{v}_\parallel^2 \left(\frac{1}{-4 \Omega_\kappa} \right) \left\{\frac{e^{-\hat{v}_\parallel^2}}{\hat{v}_\parallel^2 -\Omega/4\Omega_\kappa} \right\} + \frac{1}{\sqrt{\pi}} G \int\limits_{-\infty}^{+\infty} d \hat{v}_\parallel \left(\frac{1}{-4 \Omega_\kappa} \right) \left\{\frac{e^{-\hat{v}_\parallel^2}}{\hat{v}_\parallel^2 -\Omega/4\Omega_\kappa} \right\}, \label{eq:Q_res_intermediate_step_1}
\end{align}

where 
\begin{equation}
	G \equiv [\tau \Omega + 1 - \eta_\mathrm{i}/2] \Gamma_0(b) + \eta_\mathrm{i} b(\Gamma_1(b)-\Gamma_0(b))
\end{equation}

By introducing relabelling $\xi^2 \equiv \Omega/(4\Omega_\kappa)$, Eq.~\ref{eq:Q_res_intermediate_step_1} can be written as
\begin{align}
	\tau Q &= \frac{1}{\sqrt{\pi}} \eta_\mathrm{i} \Gamma_0(b) \left(\frac{1}{-4 \Omega_\kappa} \right) \int\limits_{-\infty}^{+\infty} d \hat{v}_\parallel \hat{v}_\parallel^2 \left\{\frac{e^{-\hat{v}_\parallel^2}}{\hat{v}_\parallel^2 - \xi^2} \right\} + \frac{1}{\sqrt{\pi}} G \left(\frac{1}{-4 \Omega_\kappa} \right) \int_{-\infty}^{+\infty} d \hat{v}_\parallel \left\{\frac{e^{-\hat{v}_\parallel^2}}{\hat{v}_\parallel^2 - \xi^2} \right\}.
\end{align}

Now utilizing the earlier defined Faddeeva function $W(\xi) \equiv \frac{1}{i\sqrt{\pi}} Z(\xi)$, 
\begin{align}
	\frac{Z(\xi)}{\xi} &= \frac{1}{\sqrt{\pi}} \int\limits_{-\infty}^{+\infty} d\hat{v}_\parallel \frac{e^{-\hat{v}_\parallel^2}}{\hat{v}_\parallel^2-\xi^2} 
\end{align}

and similarly 
\begin{align}
	\frac{1}{\sqrt{\pi}} \int\limits_{-\infty}^{+\infty} d \hat{v}_\parallel \hat{v}_\parallel^2 \left\{\frac{e^{-\hat{v}_\parallel^2}}{\hat{v}_\parallel^2 - \xi^2} \right\} &= 	\frac{1}{\sqrt{\pi}} \int\limits_{-\infty}^{+\infty} d \hat{v}_\parallel (\hat{v}_\parallel^2 - \xi^2 + \xi^2) \left\{\frac{e^{-\hat{v}_\parallel^2}}{\hat{v}_\parallel^2 - \xi^2} \right\} \nonumber \\
																					     &= 1 + \xi Z(\xi) \label{eq:Q_res_intermediate_step_2}
\end{align}

Reinserting Eq.~\ref{eq:Q_res_intermediate_step_2} back into the full expression, the ion integral reduces to 
\begin{align}
\tau Q &= \left[\left(\tau \Omega + 1 - \frac{\eta_\mathrm{i}}{2} \right) \Gamma_0(b) + \eta_\mathrm{i} b (\Gamma_1(b) - \Gamma_0(b)) \right] D_0 + \eta_\mathrm{i} \Gamma_0 D_1 \label{eq:Q_resonant}
\end{align}

with
\begin{align}
	D_0 &= -\frac{1}{4 \Omega_\kappa} \frac{Z(\xi)}{\xi},\\
	D_1 &= -\frac{1}{4 \Omega_\kappa} [1 + \xi Z(\xi)].
\end{align}

This final result is identical to Eq.~(18) of Ref.~\cite{Cheng82a} and can be directly inserted into the KBM equation Eq.~(\ref{eq:dimensionless_kbm_eqn_1}) for evaluation. Alternatively, the full unapproximated integral, Eq.~(\ref{eq:Q_unapproximated}), can be numerically evaluated from the outset by utilizing Gauss-Laguerre and Gauss-Hermite quadrature schemes for the velocity-space integration \cite{Hirose95}; results from both analytical and numerical evaluations are presented in section~\ref{sect:results}. Another analytical and higher-fidelity form of $Q$ (without the simplifying approximation $\hat{v}_\perp^2 = 2 \hat{v}_\parallel^2$ in the denominator of $Q$) has been derived for ITG theory in Ref.~\cite{Zocco18a}; this result has been found to agree exactly with the full unapproximated integral Eq.~(\ref{eq:Q_unapproximated}) when applied to a circular tokamak (not shown here). Further refinement is required regarding the numerical treatment of this latter analytical approach in order to reduce its computation time and expand its usage to stellarator geometry.  


\newpage

\subsection{KBM eigenvalue problem} \label{sect:kbm_eigenvalue_problem}

The dimensionless KBM eigenmode equation (\ref{eq:dimensionless_kbm_eqn_1}) is now reformulated as an eigenvalue problem amenable to straightforward numerical evaluation. The left-hand side of Eq.~(\ref{eq:dimensionless_kbm_eqn_1}) is rewritten as follows
\begin{equation}
	\frac{1}{\hat{J} \hat{B}} \frac{\partial}{\partial \theta} \left( \frac{g^{yy}}{\hat{J}\hat{B}^3} \frac{\partial}{\partial \theta} \Phi \right) = \frac{1}{\hat{J} \hat{B}} \left[\frac{\partial}{\partial \theta}\left(\frac{g^{yy}}{\hat{J}\hat{B}^3} \right)\frac{\partial}{\partial \theta} + \frac{g^{yy}}{\hat{J}\hat{B}^3} \frac{\partial^2}{\partial \theta^2} \right] \Phi \equiv \mathcal{L} \Phi, 
\end{equation}

where $\mathcal{L}$ is an operator acting on $\Phi$. Similarly, for the right-hand side,
\begin{equation}
	 \Omega^2 \frac{\beta_\mathrm{i}\tau^2}{2\epsilon_n^2} \left\{ \frac{Q - \alpha_\mathrm{0,i}}{1 + \tau - \tau Q} \alpha_\mathrm{0,e} - \alpha_\mathrm{1,e} \frac{2\Omega_\kappa}{\Omega} \right\} \Phi \equiv \mathcal{R} \Phi,
\end{equation}

such that the KBM equation can be written compactly as
\begin{equation}
	(\mathcal{L} - \mathcal{R}) \Phi \equiv \mathcal{M} \Phi = 0.	
\end{equation}

Given the nonlinear dependence of $\mathcal{M}$ on $\Omega$, this system can be treated as a nonlinear eigenvalue problem $\mathcal{M}(\Omega)\Phi = 0$, where one seeks to calculate the eigenvalue-eigenvector pairs $(\Omega, \Phi)$. As will be shown, the eigenvalues $\Omega$ are obtained first, and are then utilized to obtain the eigenvectors $\Phi$.  

To solve the system of linear equations represented by $\mathcal{M}(\Omega)\Phi = 0$, non-trivial solutions for the eigenvector $\Phi$ can be obtained if

\begin{equation}
	\mathrm{Det}[\mathcal{M}(\Omega)] = 0. \label{eq:detM_zero}
\end{equation}

Thus, eigenvalues $\Omega$ are sought that make $\mathcal{M}(\Omega)$ singular. In practice, determinant calculations can be susceptible to significant numerical instability, producing large-magnitude determinants even for singular matrices. As such, a more reliable method is to use condition numbers, where a singular matrix $\mathcal{M}(\Omega)$ has condition number $\mathrm{Cond}[\mathcal{M}(\Omega)]=\infty$. This is the approach applied here. Note that although $\mathrm{Cond}[\mathcal{M}(\Omega)]=\infty$ is not strictly achieved, acceptable solutions -- justified by the quality of the results obtained \textit{a posteriori} -- can be obtained by identifying $\Omega$ that maximize $|\mathrm{Cond}[\mathcal{M}(\Omega)]|$.

The procedure employed here is as follows. To obtain eigenvalues $\Omega$ that satisfy Eq.~(\ref{eq:detM_zero}), $|\mathrm{Cond}[\mathcal{M}(\Omega)]|$ is evaluated at each point on a chosen eigenvalue grid (Re $\Omega$, Im $\Omega$), where the solutions $\Omega$ that maximize $|\mathrm{Cond}[\mathcal{M}(\Omega)]|$  satisfy 

\begin{align}
	\frac{d}{d\Omega} |\mathrm{Cond}[\mathcal{M}(\Omega)]| &= 0,\\
	\frac{d^2}{d\Omega^2} |\mathrm{Cond}[\mathcal{M}(\Omega)]| &< 0,
\end{align}

where the second condition ensures $\Omega$ is at a local maximum. A new grid is then constructed in the vicinity of this previously found $\Omega$, where the mesh is made increasingly fine-grained. The maximum value of $|\mathrm{Cond}[\mathcal{M}(\Omega)]|$ near the eigenvalue solution increases with successive grid refinement, as the calculation is able to converge increasingly to the true solution. These iterations continue until a convergence criterion is met between consecutive eigenvalue results $\Omega$ (e.g., until $|\Omega_n - \Omega_{n-1}| < 10^{-10}$). Thus, although $\mathrm{Cond}[\mathcal{M}(\Omega)]=\infty$ is not reached in practice, one can get arbitrarily close to the true solution of $\Omega$ with continuous grid refinement, where each iteration allows for higher precision of the resulting $\Omega$.

The converged results for $\Omega$ are the eigenvalue solutions to the simplified KBM equation. Once these eigenvalues are obtained, all terms in $\mathcal{M}$ are known. To obtain the corresponding eigenvector $\Phi$, one can exploit the fact that the nonlinear eigenvalue problem $\mathcal{M}(\Omega) \Phi = 0$ is a special case of a simpler linear eigenvalue problem, which has eigenvector solutions $\Phi$ and eigenvalues $\lambda_{\mathcal{M}(\Omega)}$, 

\begin{equation}
	\mathcal{M}(\Omega) \Phi = \lambda_{\mathcal{M}(\Omega)} \Phi \hspace{0.5cm} \text{(for $\lambda_{\mathcal{M}(\Omega)}=0$)},
\end{equation}

such that the solution $\Phi$ of the nonlinear eigenvalue problem is identical to the solution $\Phi$ of the linear eigenvalue problem with eigenvalue $\lambda_{\mathcal{M}(\Omega)}=0$. This linear eigenvalue problem can be solved numerically and involves calculating the eigenvectors of $\mathcal{M}(\Omega)$, and selecting the solution with corresponding eigenvalue $\lambda_{\mathcal{M}(\Omega)} \approx 0$; this provides one with the KBM eigenfunction $\Phi^\mathrm{KBM}$.

\newpage

\section{Numerical evaluation of KBM theory} \label{sect:results}

In this section, the simplified KBM theory outlined in previous sections is numerically evaluated and compared with gyrokinetic flux-tube simulations performed with the \textsc{Gene} code \cite{Jenko00}, including cases that were first presented in Ref.~\cite{Mulholland23, Mulholland25}. The numerical implementation of this theory is referred to as the KBM eigenvalue yielder (\textsc{Key}) code. The \textsc{Key} code requires only flux-tube geometry and plasma parameters as inputs, and computes self-consistent KBM eigenvalues and eigenvectors for the eigenmode along the field line. Eigenvalues (eigenvectors) are calculated in $\mathcal{O}(1)$ ($\mathcal{O}(10)$) CPU seconds, respectively, where the computation time depends mostly on the field-line resolution. 

In the following subsections, \textsc{Key} and \textsc{Gene} results are compared for a sub-threshold KBM study in W7-X \cite{Mulholland23, Mulholland25}, followed by both high and low shear ($\hat{s} \approx 0.8$ and $\hat{s} = 0.1$) tokamak scenarios. In the stellarator case, results from \textsc{Gene} include the effects of trapped particles and parallel magnetic-field fluctuations $\delta B_\parallel$ (where the $\nabla B$-drift and curvature-drift frequencies are treated distinctly in conjunction with finite pressure gradient \cite{Aleynikova22}); these effects are absent in \textsc{Key}, and the $\nabla B$-drift frequency is replaced by the curvature-drift frequency ($\omega_{\nabla B} \to \omega_\kappa$) to partially compensate for the absence of $\delta B_\parallel$ \cite{Zocco15, Aleynikova22}. In addition, the \textsc{Gene} results for W7-X utilize a flux-tube geometry corresponding to one poloidal turn (for which convergence was achieved), where the field-line domain is extended artificially by exploiting the flux-tube parallel boundary condition (employing a finite number of $k_x$ modes) \cite{Merz08thesis}; this approach is more numerically efficient but can lead to notable deviations in stellarator geometry between the artificially extended domain and a flux tube with a higher number of poloidal turns \cite{Faber18}. Results from \textsc{Key} are obtained using flux-tube geometry with 20 poloidal turns and thus do not use artificially extended flux-tube domains. In the tokamak cases, results from \textsc{Gene} are obtained utilizing mostly the same assumptions as \textsc{Key} of neglecting effects from trapped particles and $\delta B_\parallel$ (using $\omega_{\nabla B} \to \omega_\kappa$). In the tokamak, artificially extending the flux-tube domain is equivalent to physical extensions using more poloidal turns. In all cases, Landau resonances are absent in \textsc{Key}, but remain in \textsc{Gene}.  

Results from \textsc{Key} include non-resonant (obtained using Eq.~(\ref{eq:Q_fluid})), analytic-resonant (partial toroidal resonance -- obtained using Eq.~(\ref{eq:Q_resonant})) and numeric-resonant (full toroidal resonance -- obtained using Eq.~(\ref{eq:Q_unapproximated})) eigenvalue (EV) solutions.


\subsection{Sub-threshold KBMs in Wendelstein 7-X} \label{sect:results_w7x}

In this section, \textsc{Key} is applied to an stKBM study carried out in stellarator geometry \cite{Mulholland23, Mulholland25}, using an MHD-optimized high-mirror configuration of W7-X denoted as KJM \cite{Aleynikova18}. For all $\beta$, the equilibria are varied consistently with the pressure gradient. The normalized gradients are $a/L_{T\mathrm{i}} = 3.5$, $a/L_{T\mathrm{e}} = 0$, $a/L_{n\mathrm{i}} = a/L_{n\mathrm{e}} = 1$, where $a$ is the minor radius and $L_{T\mathrm{j}} = -(d\ln T_j/dr)^{-1}$ and $L_{n\mathrm{j}} = -(d\ln n_j/dr)^{-1}$ are the respective length scales of the temperature and density for particle species $j$. 

\begin{figure}[h]
	\centering
	\includegraphics[width=12cm]{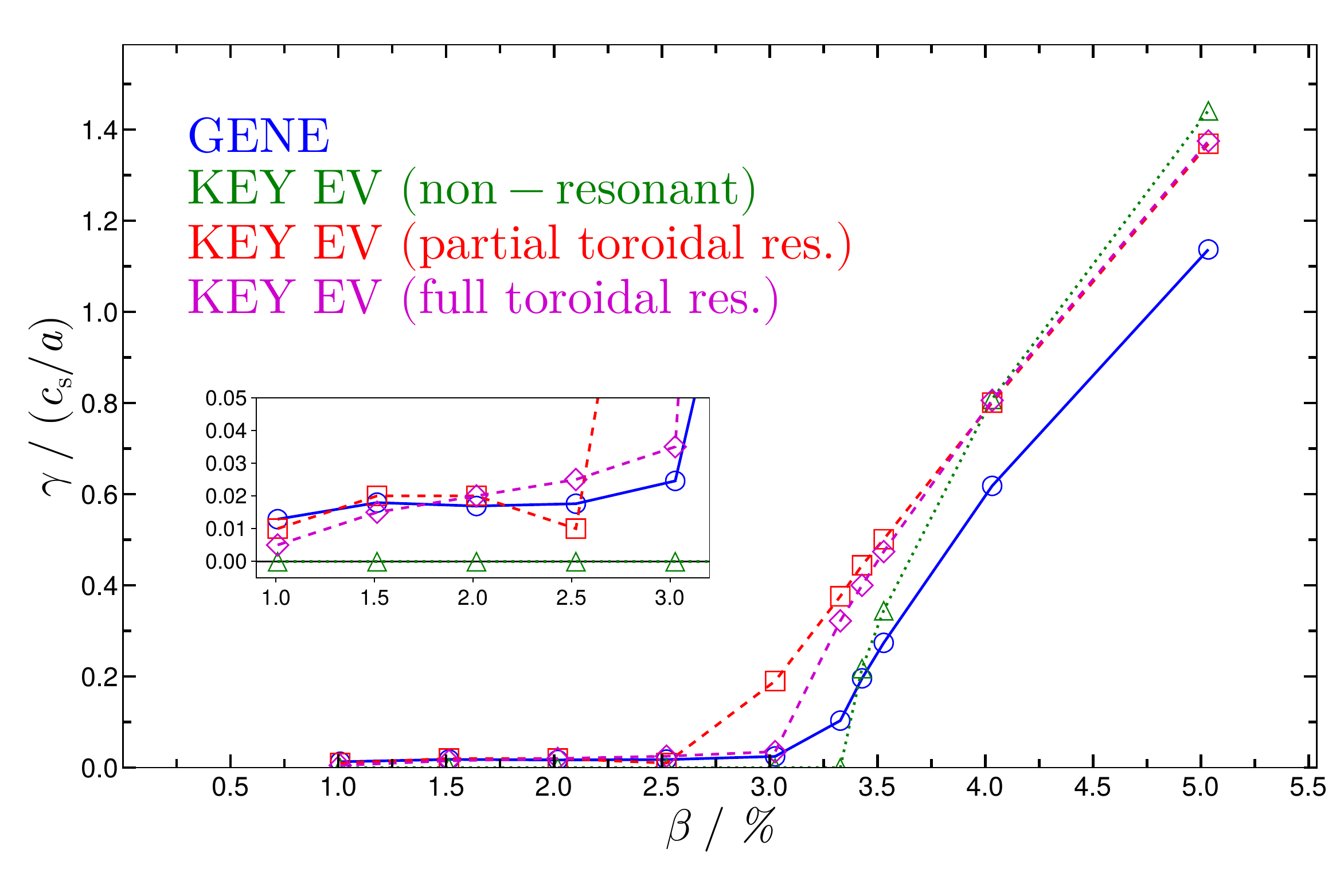}
	\includegraphics[width=12cm]{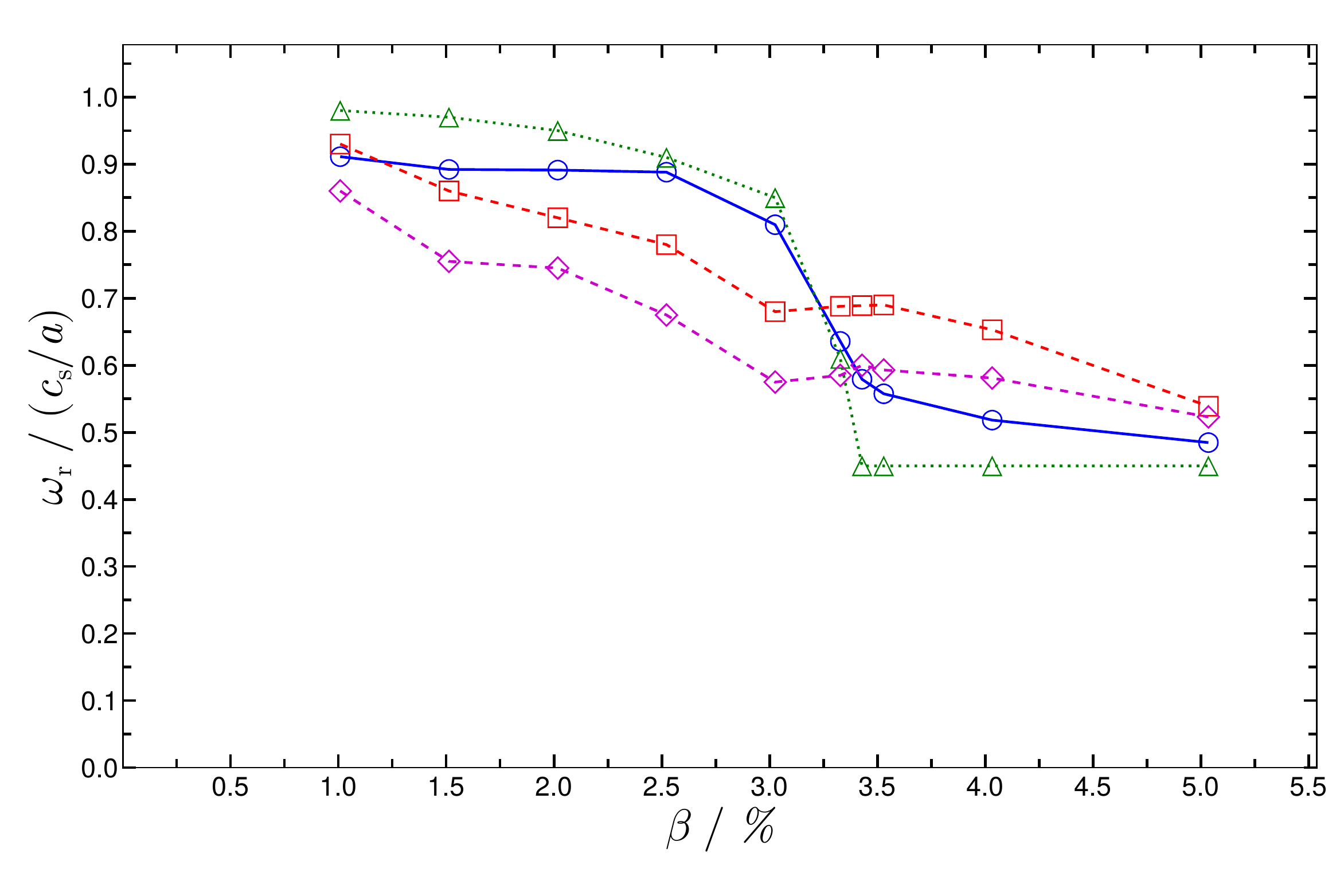}
	\caption{(st)KBM eigenvalues (growth rates $\gamma$ and real frequencies $\omega_\mathrm{r}$) with increasing $\beta$ in W7-X KJM geometry at wavenumber $k_y \rho_\mathrm{s} = 0.2$, comparing \textsc{Gene} and \textsc{Key}. \textsc{Key} curves include results from non-resonant (green triangles, dotted lines), analytic-resonant (partial toroidal resonance; red squares, dashed lines), and numeric-resonant (full toroidal resonance; magenta diamonds, dashed lines) eigenvalue (EV) solutions. Results from \textsc{Gene} (blue circles, solid lines) include effects of trapped particles and $\delta B_\parallel$, which are absent in \textsc{Key}. For the non-resonant approach, stable KBMs are obtained until reaching the predicted conventional-KBM regime at $\beta_\mathrm{crit}^\mathrm{KBM} \approx 3.3\%$, beyond which the mode grows rapidly; both resonant approaches detect low-$\beta$ destabilization of the stKBM at $\beta_\mathrm{crit}^\mathrm{stKBM} \approx 1\%$ and the mode's approximately constant growth rate until reaching $\beta_\mathrm{crit}^\mathrm{KBM} \approx 3.3\%$. All solutions return similar frequencies with $\beta$; beyond $\beta_\mathrm{crit}^\mathrm{KBM} \approx 3.3\%$, the non-resonant approach recovers the theoretically predicted frequency for strongly driven KBMs of $\omega_\mathrm{r} = \omega_\mathrm{*pi}/2$.} 
	\label{fig:stkbm_evals_w7x_kjm_key_kinetic_evp_fluid_evp_vs_gene}
\end{figure}

Figure \ref{fig:stkbm_evals_w7x_kjm_key_kinetic_evp_fluid_evp_vs_gene} shows growth rates $\gamma$ and real frequencies $\omega_\mathrm{r}$ with increasing $\beta$ for the (st)KBM from \textsc{Key} and \textsc{Gene} at normalized binormal wavenumber $k_y \rho_\mathrm{s} = 0.2$. Focus is given to low binormal wavenumbers as KBM studies in W7-X have shown that growth rates peak at $k_y \rho_\mathrm{s} \to 0$ \cite{Aleynikova18, Aleynikova22, Mulholland25}. Results from \textsc{Key} include non-resonant, analytic-resonant, and numeric-resonant EV solutions. Results from \textsc{Gene} include effects of trapped particles and $\delta B_\parallel$, which are absent in \textsc{Key}. In both \textsc{Gene} and \textsc{Key}, two distinct eigenmode clusters are obtained corresponding to KBMs and electromagnetic ITGs; focus is given here only to the KBM cluster. The stKBM, once rendered unstable at $\beta_\mathrm{crit}^\mathrm{stKBM} \approx 1\%$, is a single mode within the KBM cluster that remains subdominant ($\gamma < \gamma_\mathrm{max}$) until $\beta_\mathrm{crit}^\mathrm{KBM} \approx 3.3\%$, from which point it grows steeply with $\beta$ as a conventional KBM \cite{Mulholland23, Mulholland25}. In \textsc{Key}, the non-resonant approach finds that the mode is stable until surpassing $\beta = 3\%$, whereas the resonant approaches show instability from as low as $\beta \approx 1\%$. Thus, when accounting for ion-magnetic-drift resonances, \textsc{Key} is shown to capture both the unconventional characteristics of the stKBM at low $\beta$ as well as the more typical behavior expected from the conventional KBM at higher $\beta$. Specifically, \textsc{Key} reproduces the low-$\beta$ destabilization at $\beta_\mathrm{crit}^\mathrm{stKBM} \approx 1\%$ and the mode's approximately constant growth rate until reaching the conventional-KBM regime at $\beta_\mathrm{crit}^\mathrm{KBM} \approx 3.3\%$, where the iMHD limit is reached and the mode grows rapidly with $\beta$. All approaches return similar but distinct frequency trends with $\beta$; beyond $\beta_\mathrm{crit}^\mathrm{KBM} \approx 3.3\%$, the non-resonant results closely correspond to the theoretically predicted value for strongly driven KBMs \cite{Tang80} of $\omega_\mathrm{r} = \omega_\mathrm{*pi}/2$, where $\omega_\mathrm{*pi} = \omega_\mathrm{*i}(1+\eta_\mathrm{i})$ is the ion diamagnetic frequency associated with the ion pressure gradient. A noteworthy feature is the common frequency trend for all curves (\textsc{Gene}, non-resonant and resonant) with increasing $\beta$; the non-resonant results aid in explaining this phenomenon. For $\beta < 3\%$, where the KBM is stable in the non-resonant picture, \textsc{Key} returns a pair of real solutions that are symmetric about $\omega_\mathrm{r} = \omega_\mathrm{*pi}/2$ (only one solution is shown here); once criticality is reached at $\beta_\mathrm{crit}^\mathrm{KBM} \approx 3.3\%$, these two real solutions transition to a complex-conjugate pair with a common real part of $\omega_\mathrm{r} = \omega_\mathrm{*pi}/2$ and imaginary parts $\pm i\gamma$, corresponding to one unstable and one stable branch, respectively (only the unstable branch is shown here). Therefore, the similarities seen between the non-resonant and resonant results may indicate that the introduction of resonances mainly acts to augment the non-resonant picture, but does not drastically alter it. Regardless, accounting for the differences brought about by ion-magnetic-drift resonances -- though subtle -- is vital, given that their inclusion explains the deleterious manifestation of the stKBM and its secondary effect on ITG heat fluxes. 

\begin{figure}[h]
	\centering
	\includegraphics[width=15.5cm]{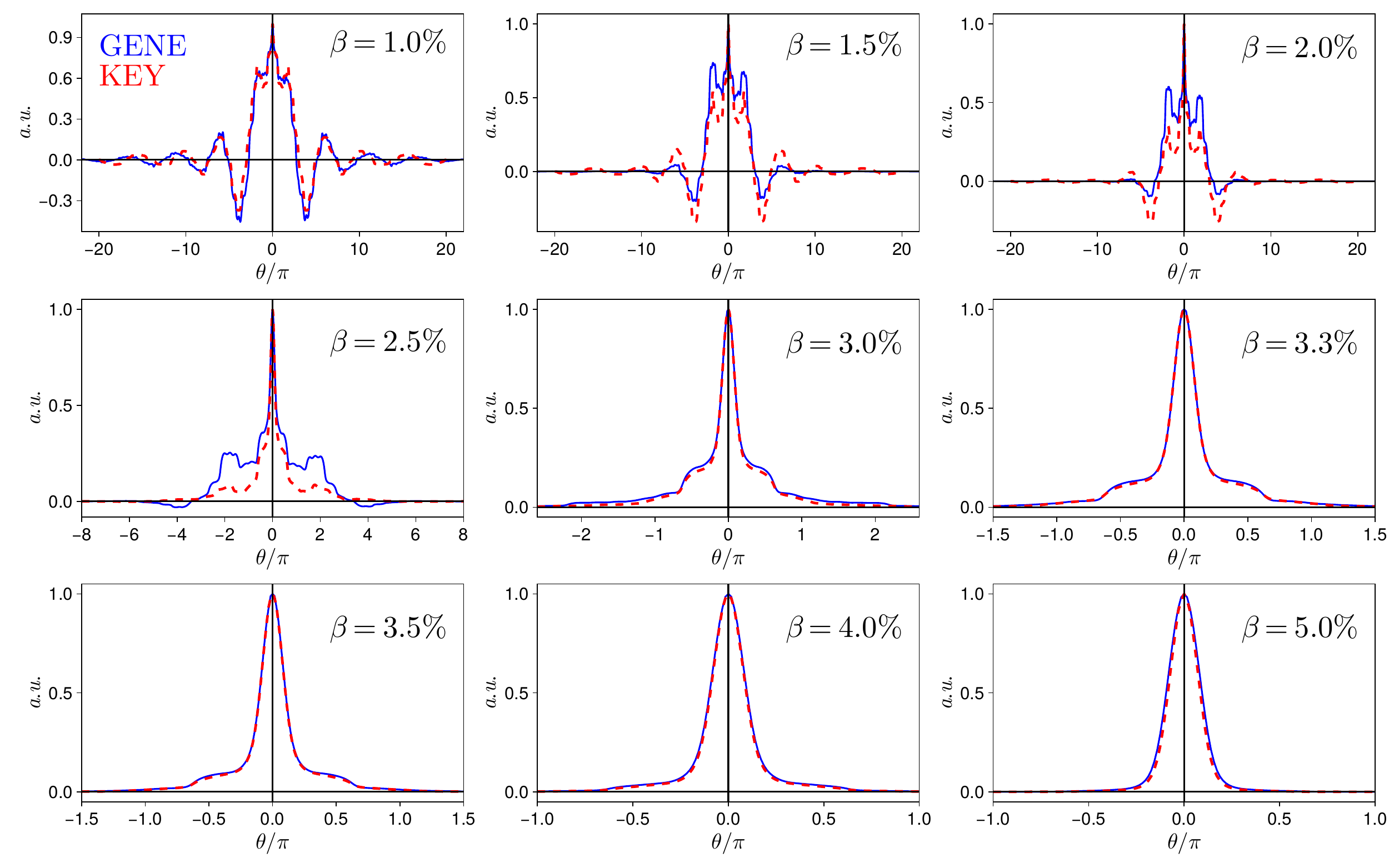}
	\caption{(st)KBM eigenfunctions $\mathrm{Re}\,\Phi$ versus ballooning angle $\theta$ at different $\beta$ in W7-X KJM geometry at wavenumber $k_y \rho_\mathrm{s} = 0.2$, comparing \textsc{Gene} (blue solid lines) and \textsc{Key} (red dashed lines), where \textsc{Key} eigenfunctions correspond to the analytic-resonant eigenvalues in Fig.~\ref{fig:stkbm_evals_w7x_kjm_key_kinetic_evp_fluid_evp_vs_gene}. Results from \textsc{Gene} include effects of trapped particles and $\delta B_\parallel$, which are absent in \textsc{Key}. \textsc{Key} captures the broad mode structure of the stKBM for $\beta \approx 1\%-3\%$ in the complex geometry of W7-X, as well as its progressive narrowing with increasing drive as it transitions to the conventional KBM for $\beta > 3\%$.} \label{fig:stkbm_efns_w7x_kjm_key_kinetic_evp_vs_gene}
\end{figure}

Figure \ref{fig:stkbm_efns_w7x_kjm_key_kinetic_evp_vs_gene} shows eigenfunctions of the real part of the perturbed electrostatic potential $\mathrm{Re}\,\Phi$ versus ballooning angle $\theta$ from \textsc{Gene} and \textsc{Key} for the stKBM ($\beta \approx 1\%-3\%$) and conventional KBM ($\beta > 3\%$) at normalized binormal wavenumber $k_y \rho_\mathrm{s} = 0.2$. Eigenfunctions from \textsc{Key} correspond to the analytic-resonant EV approach; other EV approaches (non-resonant and numeric-resonant) return similar eigenfunctions to those shown here. Eigenfunctions from both resonant approaches in \textsc{Key} have $\mathrm{Re}\,\Phi \gg \mathrm{Im}\,\Phi$, while in the non-resonant approach, $\mathrm{Im}\,\Phi = 0$. Only $\mathrm{Re}\,\Phi$ is shown here to highlight the positive-negative oscillations of the stKBM eigenfunction at $|\theta| \neq 0$ for low $\beta$ observed in both \textsc{Gene} and \textsc{Key}. \textsc{Key} is shown to capture the broad mode structure of the stKBM in the complex geometry of W7-X, as well as its progressive narrowing with increasing drive as it transitions to the conventional KBM. The shape of the stKBM eigenfunction is primarily determined by the field-line curvature $\mathcal{K}_y$ in W7-X; specifically, the high spatial variation of bad-curvature wells in W7-X gives the broad stKBM eigenfunction its modulated amplitude along the field line, where the eigenmode peaks at locations where bad-curvature wells coincide with regions of lower $g^{yy}$, where the latter provides the stabilizing influence of magnetic shear and FLR suppression. Discrepancies between the results from \textsc{Gene} and \textsc{Key} are likely attributed to differences in the underlying assumptions of each code combined with deviations in flux-tube geometry for $\theta/\pi > 1$, i.e., beyond one poloidal turn, \textsc{Gene} uses artificial extensions of the geometry whereas \textsc{Key} uses a longer flux tube of 20 poloidal turns. Improved agreement is expected if the same underlying assumptions are made and if identical flux-tube domains are employed by both codes.


\subsection{High-/low-shear tokamak} \label{sect:results_tokamak}

\textsc{Key} is now applied to a tokamak scenario adapted from the Cyclone Base Case \cite{Dimits00a}, where both nominal background magnetic shear ($\hat{s} \approx 0.8$) and a reduced shear ($\hat{s} = 0.1$) are considered. In both cases, the equilibria are varied consistently with the pressure gradient, trapped particles and $\delta B_\parallel$ are neglected, the $\nabla B$-drift frequency is replaced by the curvature-drift frequency ($\omega_{\nabla B} \to \omega_\kappa$), and the normalized gradients are $R/L_{T\mathrm{i}} = R/L_{T\mathrm{e}} = R/L_{n\mathrm{i}} = R/L_{n\mathrm{e}} = 10$, where $R$ is the major radius. 

\begin{figure}[h]
	\centering
	\includegraphics[width=12cm]{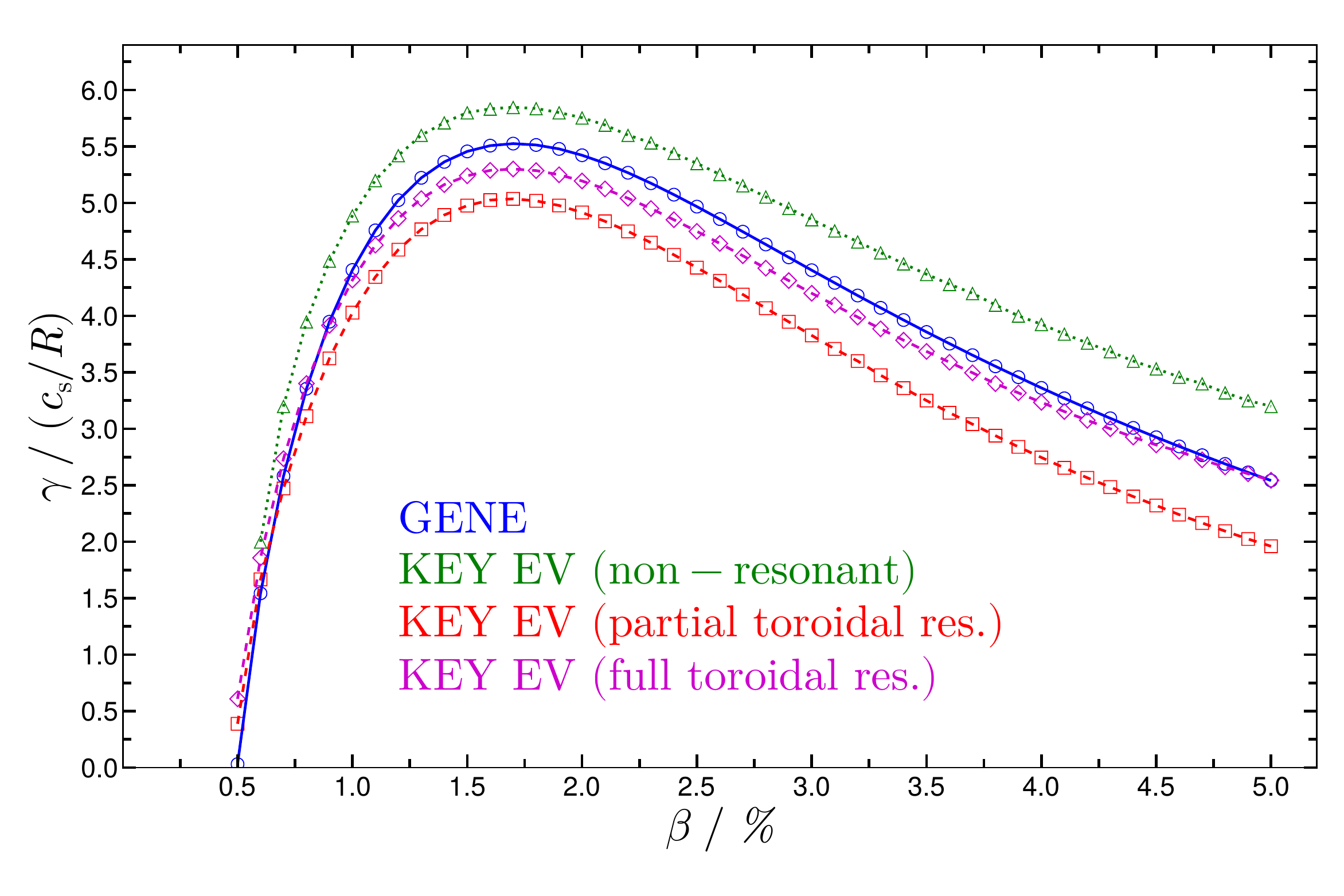}
	\includegraphics[width=12cm]{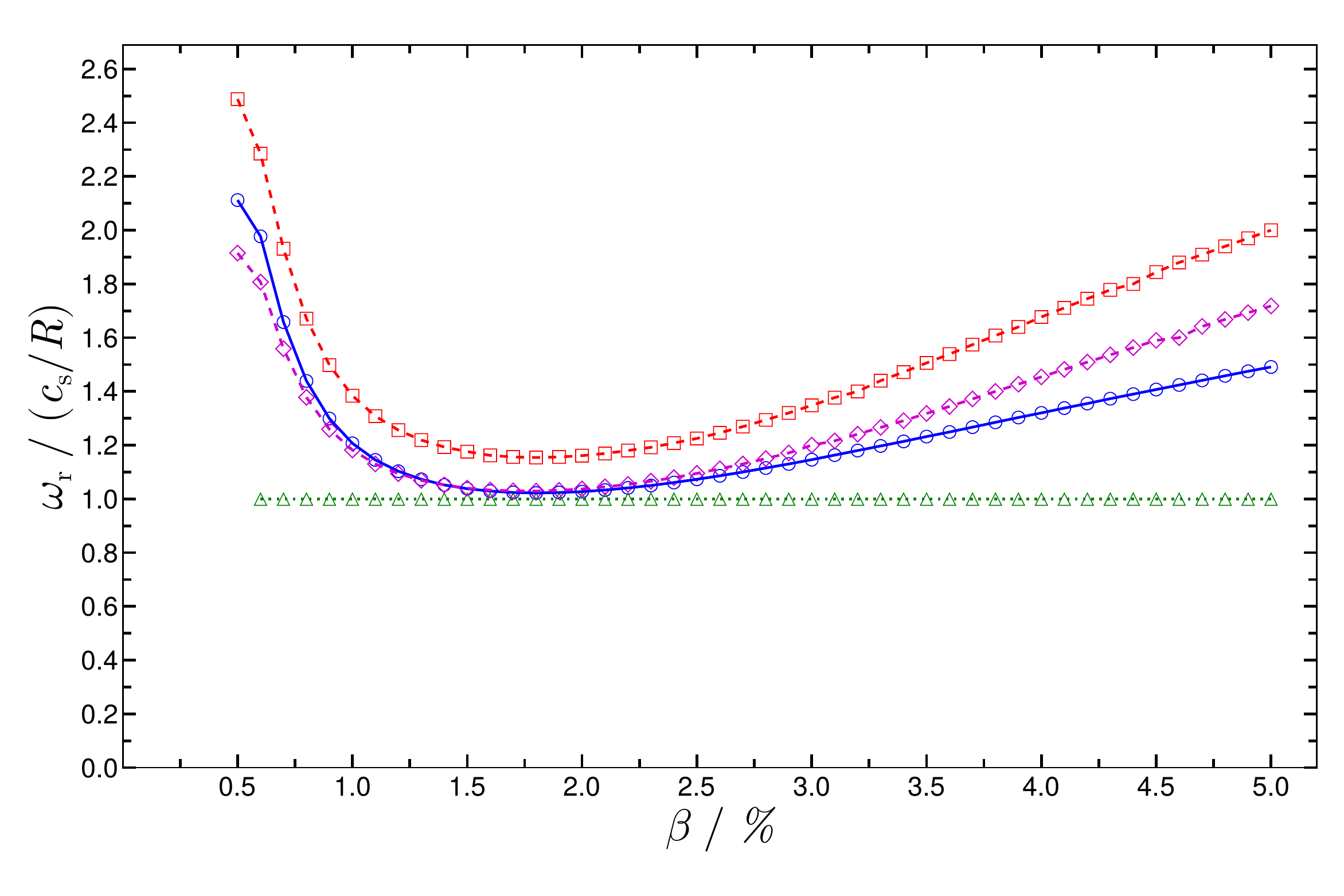}
	\caption{KBM eigenvalues (growth rates $\gamma$ and real frequencies $\omega_\mathrm{r}$) as functions of $\beta$ in tokamak geometry with $\hat{s} \approx 0.8$ at wavenumber $k_y \rho_\mathrm{s} = 0.1$, comparing \textsc{Gene} and \textsc{Key}. \textsc{Key} curves include results from non-resonant (green triangles, dotted lines), analytic-resonant (partial toroidal resonance; red squares, dashed lines), and numeric-resonant (full toroidal resonance; magenta diamonds, dashed lines) eigenvalue (EV) solutions. Growth rates from \textsc{Key} show fair but not perfect agreement with \textsc{Gene} (blue circles, solid lines); resonant results follow \textsc{Gene} more consistently than non-resonant results. For the frequencies, the resonant approaches follow the trend of \textsc{Gene}, while the non-resonant approach recovers the theoretically predicted value for strongly driven KBMs of $\omega_\mathrm{r} = \omega_\mathrm{*pi}/2$. Resonant results from \textsc{Key} recover $\beta_\mathrm{crit}^\mathrm{KBM} \approx 0.5\%$, in agreement with \textsc{Gene}. The increasing stabilization seen for $\beta > 2\%$ is well captured in \textsc{Key}.}
	\label{fig:kbm_evals_salpha_shat_0.796_key_evp_kinetic_vs_gene}
\end{figure}

Figure \ref{fig:kbm_evals_salpha_shat_0.796_key_evp_kinetic_vs_gene} shows growth rates $\gamma$ and real frequencies $\omega_\mathrm{r}$ with increasing $\beta$ from \textsc{Key} and \textsc{Gene} for the KBM with $\hat{s} \approx 0.8$ at normalized binormal wavenumber $k_y \rho_\mathrm{s} = 0.1$. Results from \textsc{Key} include non-resonant, analytic-resonant, and numeric-resonant EV solutions. In all approaches, growth rates from \textsc{Key} show qualitative agreement with \textsc{Gene} (blue circles, solid lines) in terms of capturing the overall trend with increasing $\beta$. For the frequencies, the resonant approaches approximately follow the trend of \textsc{Gene}, while the non-resonant approach recovers the theoretically predicted value for strongly driven KBMs of $\omega_\mathrm{r} = \omega_\mathrm{*pi}/2$. In alignment with this, frequencies from the resonant approaches and \textsc{Gene} tend toward this strong-drive frequency $\omega_\mathrm{r} \to \omega_\mathrm{*pi}/2$ as the growth rate peaks at $\beta \approx 1.5\%-2\%$. Both resonant approaches in \textsc{Key} obtain the same destabilization threshold as \textsc{Gene} of $\beta_\mathrm{crit}^\mathrm{KBM} \approx 0.5\%$, whereas the non-resonant approach overestimates this threshold to be $\beta_\mathrm{crit}^\mathrm{KBM} \approx 0.6\%$. The distinct feature of increasing stabilization of the KBM at high $\beta > 2\%$ is well captured in all approaches. 

\begin{figure}[h]
	\centering
	\includegraphics[width=18cm]{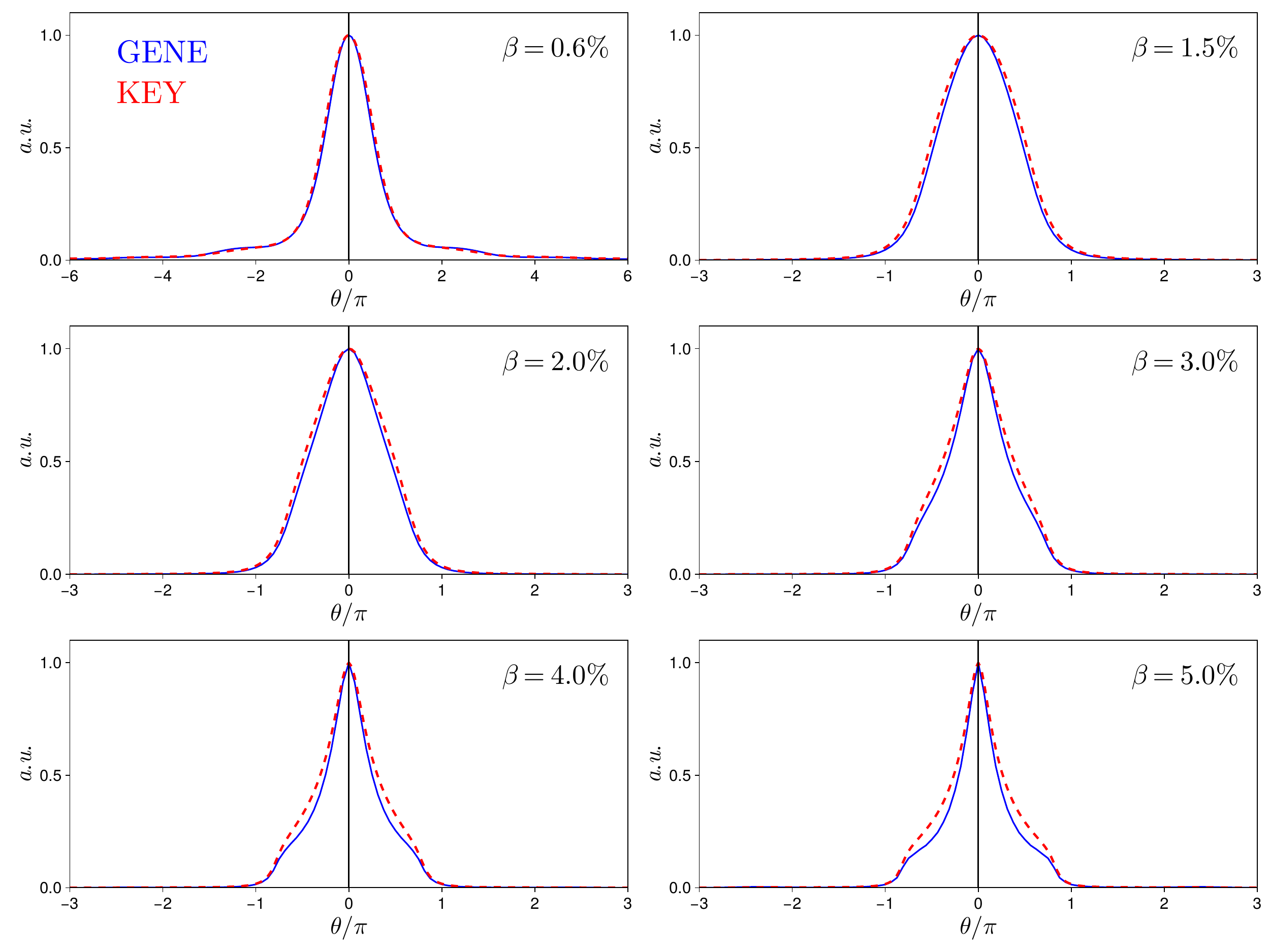}
	\caption{KBM eigenfunctions $|\Phi|$ versus ballooning angle $\theta$ at different $\beta$ in tokamak geometry with $\hat{s} \approx 0.8$ at wavenumber $k_y \rho_\mathrm{s} = 0.1$, comparing \textsc{Gene} (blue solid lines) with \textsc{Key} (red dashed lines), where \textsc{Key} eigenfunctions correspond to the numeric-resonant eigenvalues in Fig.~\ref{fig:kbm_evals_salpha_shat_0.796_key_evp_kinetic_vs_gene}. Good agreement is found for all $\beta$.}
	\label{fig:kbm_efns_salpha_shat_0.796_key_evp_kinetic_vs_gene}
\end{figure}

Figure \ref{fig:kbm_efns_salpha_shat_0.796_key_evp_kinetic_vs_gene} shows eigenfunctions of the perturbed electrostatic potential $\Phi$ with ballooning angle $\theta$ from \textsc{Gene} and \textsc{Key} for the KBM with $\hat{s} \approx 0.8$ at normalized binormal wavenumber $k_y \rho_\mathrm{s} = 0.1$ at different $\beta$; these eigenfunctions correspond to the numeric-resonant eigenvalues shown in Fig.~\ref{fig:kbm_evals_salpha_shat_0.796_key_evp_kinetic_vs_gene}. Near marginality, both codes obtain a centrally peaked eigenfunction with decreasing amplitude at higher $\theta$. With increasing $\beta \approx 1\%$, the mode narrows slightly as the growth rate increases sharply. At higher $\beta > 2\%$, the growth rate reduces and  the eigenfunction broadens into a bell shape, present in both \textsc{Gene} and \textsc{Key}. Good agreement is found between both codes for all $\beta$. 

\begin{figure}[h]
	\centering
	\includegraphics[width=12cm]{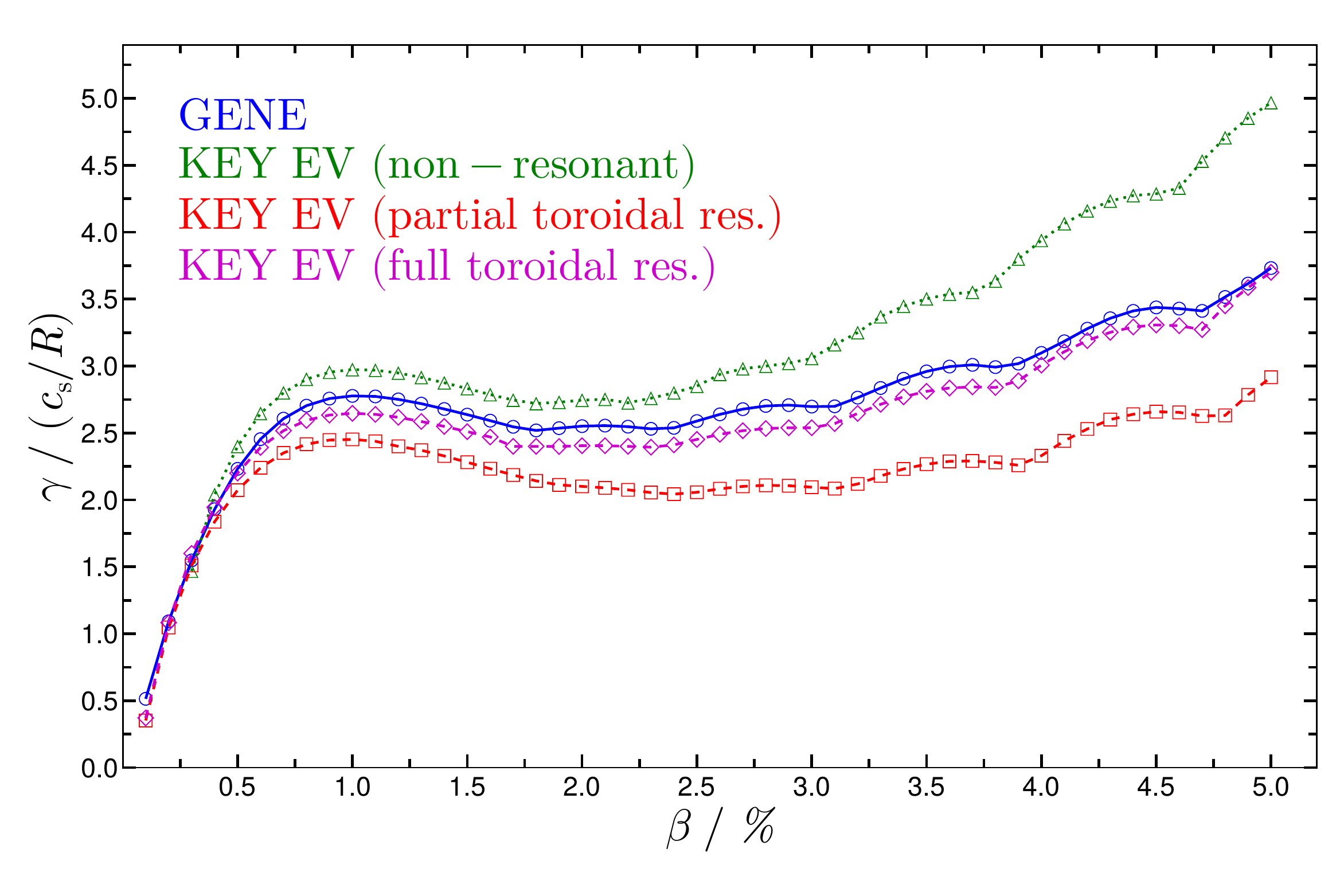}
	\includegraphics[width=12cm]{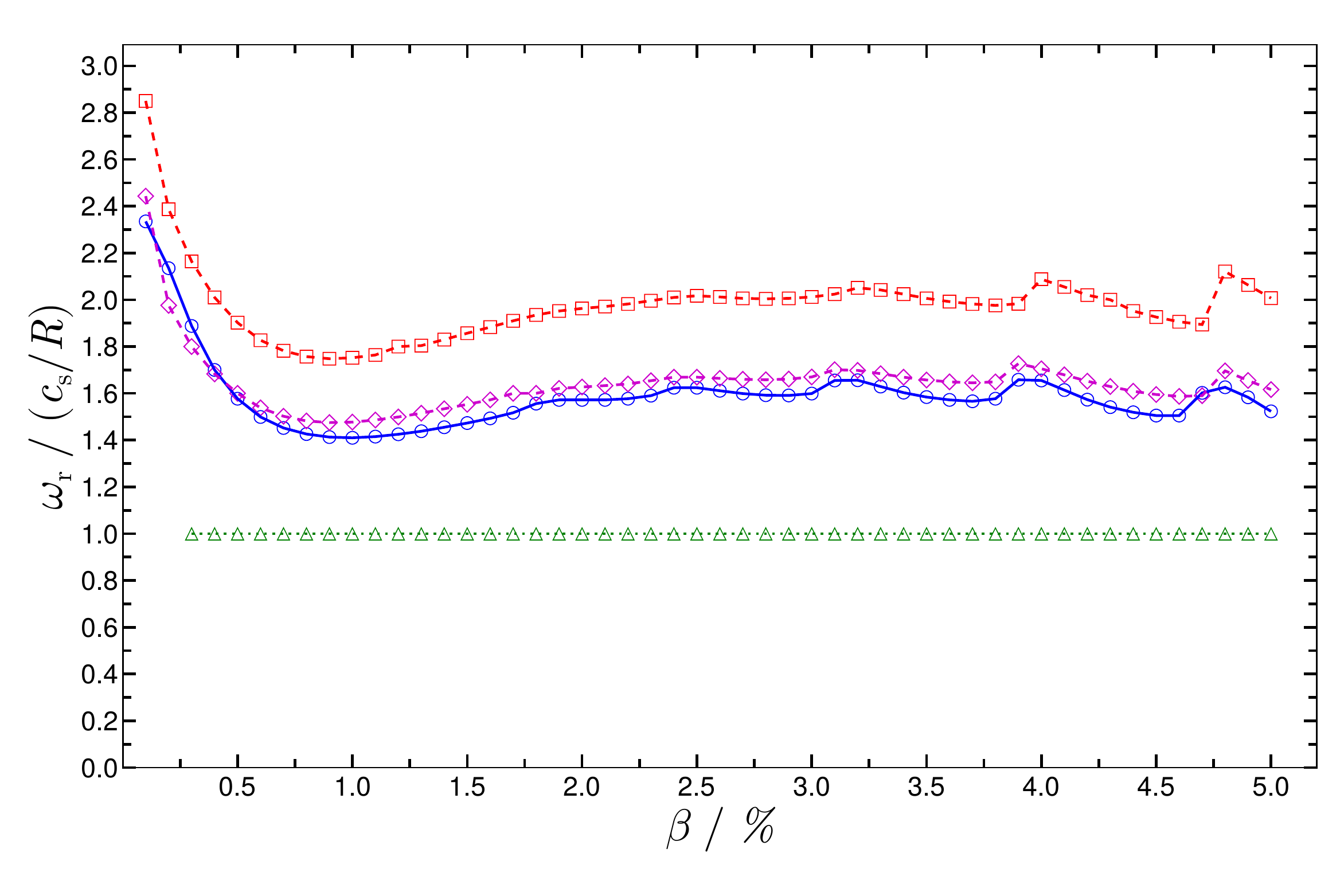}
	\caption{KBM eigenvalues (growth rates $\gamma$ and real frequencies $\omega_\mathrm{r}$) as functions of $\beta$ in tokamak geometry with $\hat{s} \approx 0.1$ at wavenumber $k_y \rho_\mathrm{s} = 0.1$, comparing \textsc{Gene} and \textsc{Key}. \textsc{Key} curves include results from non-resonant (green triangles, dotted lines), analytic-resonant (partial toroidal resonance; red squares, dashed lines), and numeric-resonant (full toroidal resonance; magenta diamonds, dashed lines) eigenvalue (EV) solutions. Growth rates from \textsc{Key} show qualitative agreememnt with \textsc{Gene} (blue circles, solid lines); resonant results follow \textsc{Gene} more consistently than non-resonant results. For the frequencies, the resonant approaches follow the trend of \textsc{Gene}, while the non-resonant approach recovers the theoretically predicted value for strongly driven KBMs of $\omega_\mathrm{r} = \omega_\mathrm{*pi}/2$. \textsc{Key} slightly overestimates $\beta_\mathrm{crit}^\mathrm{KBM}$ at $0.1\%-0.3\%$, compared with \textsc{Gene}'s $\beta_\mathrm{crit}^\mathrm{KBM} \approx 0.05\%$. The modulating $\gamma$ with increasing $\beta$ is well captured in \textsc{Key}.}
	\label{fig:kbm_evals_salpha_shat_0.1_key_evp_kinetic_vs_gene}
\end{figure}

Figure \ref{fig:kbm_evals_salpha_shat_0.1_key_evp_kinetic_vs_gene} shows growth rates $\gamma$ and real frequencies $\omega_\mathrm{r}$ with increasing $\beta$ from \textsc{Key} and \textsc{Gene} for the KBM with $\hat{s} \approx 0.1$ at normalized binormal wavenumber $k_y \rho_\mathrm{s} = 0.1$. Results from \textsc{Key} include non-resonant, analytic-resonant, and numeric-resonant EV solutions. In all approaches, \textsc{Key} captures the modulating but generally increasing $\gamma$ with $\beta$; this modulation is caused by the interchanging of eigenmodes for the dominant position (largest $\gamma$) as $\beta$ increases, where the results shown here correspond only to the dominant mode at any given $\beta$. It is noteworthy to mention that \textsc{Key}, like \textsc{Gene}, can recover subdominant modes. The modulating behavior observed here resembles a previous study of KBMs in a low-shear tokamak, reported in Ref.~\cite{Hirose95}. For the frequencies, the resonant approaches approximately follow the trend of \textsc{Gene}, while the non-resonant approach recovers the theoretically predicted value for strongly driven KBMs of $\omega_\mathrm{r} = \omega_\mathrm{*pi}/2$. Agreement between \textsc{Gene} and \textsc{Key} improves with the increasing physics fidelity of each approach (i.e., non-resonant $\to$ partial toroidal resonance $\to$ full toroidal resonance). The non-resonant approach overestimates the growth rate for $\beta > 0.4\%$, and deviations increase with $\beta$. The analytic-resonant approach improves upon the former, but consistently underestimates the growth rate for all $\beta$. Lastly, the numeric-resonant approach obtains the best agreement with \textsc{Gene}, though it slightly underestimates the growth rate for $\beta > 0.5\%$. All approaches in \textsc{Key} slightly overestimate the destabilization threshold to be $\beta_\mathrm{crit}^\mathrm{KBM} \approx 0.1\%-0.3\%$, compared to \textsc{Gene}'s estimated extrapolated threshold of $\beta_\mathrm{crit}^\mathrm{KBM} \approx 0.05\%$. The relative difference between these thresholds is substantial, but is amplified by the very low $\beta$ here; in terms of absolute differences, these deviations are minor. In terms of comparing machine performance, \textsc{Key} accurately detects the notable differences in destabilization thresholds and growth-rate trends observed in the high- and low-shear tokamaks presented here. 

\begin{figure}[h]
	\centering
	\includegraphics[width=18cm]{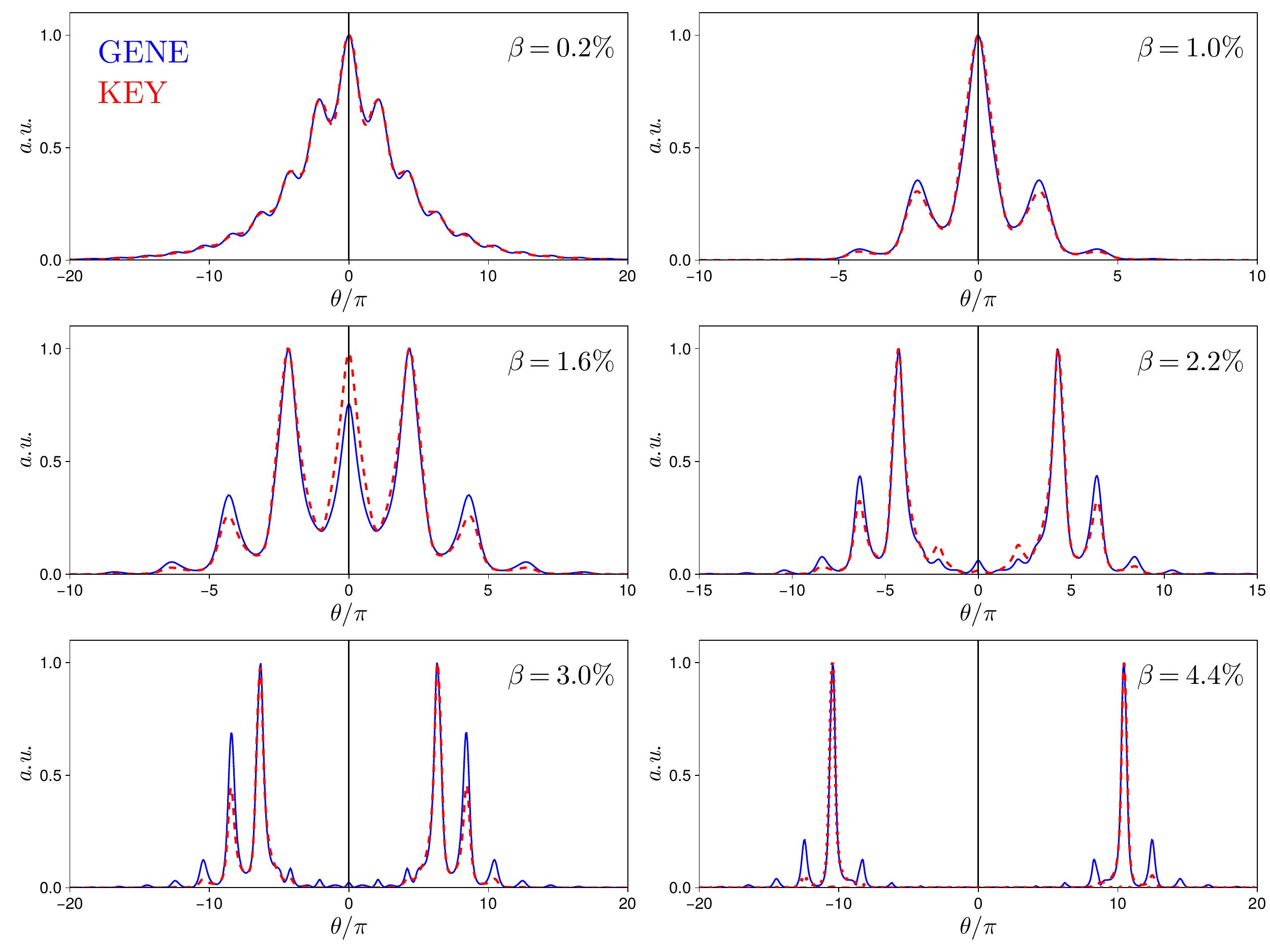}
	\caption{KBM eigenfunctions $|\Phi|$ versus ballooning angle $\theta$ at different $\beta$ in tokamak geometry with $\hat{s} \approx 0.1$ at wavenumber $k_y \rho_\mathrm{s} = 0.1$, comparing \textsc{Gene} (blue solid lines) and \textsc{Key} (red dashed lines), where \textsc{Key} eigenfunctions correspond to the numeric-resonant eigenvalues in Fig.~\ref{fig:kbm_evals_salpha_shat_0.1_key_evp_kinetic_vs_gene}. Good agreement is found up to $\beta \approx 1\%$, beyond which some discrepancies become manifest, but substantial agreement persists. At $\beta \geq 3.5\%$, \textsc{Gene} produces a symmetric eigenfunction while \textsc{Key} obtains two identical modes mirrored across $\theta=0$ (dashed and dotted red lines); the combination of these mirrored modes agrees with the single mode from \textsc{Gene}. Results obtained for $\beta > 1.7\%$ likely correspond to higher-excitation states of the KBM whose substantial amplitude at high $|\theta|$ is owed to the low background magnetic shear of this configuration and the influence of increasing $\alpha_\mathrm{MHD}$ (see text for details).}
	\label{fig:kbm_efns_salpha_shat_0.1_key_evp_kinetic_vs_gene}
\end{figure}

Figure \ref{fig:kbm_efns_salpha_shat_0.1_key_evp_kinetic_vs_gene} shows eigenfunctions of the perturbed electrostatic potential $\Phi$ with ballooning angle $\theta$ from \textsc{Gene} and \textsc{Key} for the KBM with $\hat{s} = 0.1$ at normalized binormal wavenumber $k_y \rho_\mathrm{s} = 0.1$ at different $\beta$; these eigenfunctions correspond to the numeric-resonant eigenvalues shown in Fig.~\ref{fig:kbm_evals_salpha_shat_0.1_key_evp_kinetic_vs_gene}. Near marginality, both codes obtain very broad eigenfunctions that gradually decrease in amplitude up to $|\theta| \approx 20 \pi$, where the mode has an oscillatory structure. With increasing $\beta \approx 1\%$, the amplitude of the mode reduces strongly for $|\theta| > 5 \pi$, and some oscillatory structure remains near for $|\theta| < 5\pi$. At $\beta \approx 1.6\%$, the shoulders of the mode increase in amplitude. For higher $\beta$, peaks become more prominent away from $\theta=0$, far exceeding the central-peak value. At $\beta \approx 2.2\%$, the central peak has almost fully disappeared while substantial peaks grow at $|\theta| \approx 5\pi - 10\pi$. This trend continues for higher $\beta > 3\%$, where the only remaining peaks occur away from $\theta=0$, and which move to higher $|\theta|$ with increasing $\beta$. For $\beta > 3\%$, \textsc{Gene} produces a single symmetric eigenfunction, while \textsc{Key} obtains two identical modes which are mirrored across $\theta=0$ (shown in dashed and dotted red lines); the combination of these mirrored modes resembles the single mode from \textsc{Gene}. Such mirror-mode pairs with a shared eigenvalue have been found in previous KBM studies using high-fidelity eigenvalue solver approaches \cite{McKinney21, Mulholland23, Mulholland25}. These mirror-mode pairs may correspond to two finite-ballooning angle modes having radial wavenumbers $+k_x$ and $-k_x$, where the ballooning angle can be expressed as $\theta_b = -k_x/(\hat{s}k_y)$. The results obtained here for higher $\beta > 1.6\%$ are atypical, as KBMs are generally expected to have centrally peaked eigenfunctions with moderate-to-low amplitude at $\theta \neq 0$; these modes at high $\beta$ are presumably higher-excitation states \cite{Pueschel19} of the KBM. The shape of these eigenfunctions is primarily determined by the field-line curvature $\mathcal{K}_y$ and the influence of magnetic shear manifest in $g^{yy}$ (where the latter is also related to FLR damping). In particular, in this circular tokamak geometry, the bad-curvature regions that drive the eigenmode are located at $\theta = 2\pi m$, where $m$ corresponds to positive and negative integers; the eigenfunction amplitude peaks at these locations if the stabilizing influence of $g^{yy}$ is not too strong there. The variation of $g^{yy} = 1 + [\hat{s}\theta - \alpha_\mathrm{MHD} \sin(\theta)]^2$ with increasing $\alpha_\mathrm{MHD}$ causes the relocation of the eigenfuction peaks to higher $\theta$ with increasing $\beta$. It is noteworthy that KBM eigenfunctions from \textsc{Key} and \textsc{Gene} can also have tearing parity (odd parity in $\Phi$, even parity in $A_\parallel)$, which are less common but sometimes manifest (see Ref.~\cite{McKinney21}). Despite the unconventional nature of these highly excited eigenmodes, \textsc{Key} manages to obtain good agreement with \textsc{Gene}. 

\newpage

\subsection{Analysis and physical interpretation of (st)KBMs} \label{sect:results_discussion}

The results in the previous section show that \textsc{Key} accurately captures important features of KBMs in both the stellarator and the tokamak; notably, \textsc{Key} reveals that wave-particle-resonances from passing ions subject to the magnetic-curvature drift (i.e., toroidal resonances) must be accounted for in order to detect the presence of stKBMs. In the tokamak, \textsc{Key} qualitatively captures the behavior of conventional KBMs in both high- and low-magnetic-shear scenarios, approximately reproducing the eigenvalue trends and eigenfunctions in each case. 

In this section, the results from \textsc{Key} are discussed further with a focus on the distinctions between KBMs in the stellarator and the tokamak; specifically, the presence of stKBMs in W7-X and their absence in the tokamak cases studied here. A physical interpretation is provided for the mechanism of the stKBM and the conditions required for its existence. 

Certain conditions are known to be necessary in order for a device to host an stKBM, while other conditions remain speculative and require further scrutiny; assessing the latter is planned for future work. The known conditions pertain both to the plasma itself and to the magnetic geometry in question. Regarding the plasma, a finite ion-temperature gradient ($\eta_\mathrm{i} \neq 0$) is essential in order for the resonant threshold of destabilization to be below the non-resonant threshold; $\eta_\mathrm{i} = 0$ leads to both approaches sharing a common $\beta_\mathrm{crit}^\mathrm{KBM}$ at the non-resonant threshold, and increasing $\eta_\mathrm{i}$ leads to an incremental reduction in the resonant threshold relative to the non-resonant counterpart \cite{Cheng82a}. It follows that another requirement for stKBM manifestation is the presence of ion-magnetic-drift resonances; a purely non-resonant model is insufficient to capture important details of the KBM near marginality, such that the low-$\beta$ destabilization of the stKBM would be missed. However, it should be noted that the non-resonant approach used here is capable of producing KBM eigenfunctions (not shown here) that are very similar to those obtained in the resonant approaches in Fig.~\ref{fig:stkbm_efns_w7x_kjm_key_kinetic_evp_vs_gene}, even in the stKBM region of $\beta \approx 1\%-3\%$. This indicates that stKBMs arise from a combination of non-resonant and kinetic physics: the non-resonant picture largely captures the eigenfunction shape and approximate eigenvalue trends, while the resonant picture allows access to more accurate eigenvalues at lower drive near marginality, and thus determines the actual stability threshold of the (st)KBM, i.e., whether unstable stKBMs are present, where $\beta_\mathrm{crit}^\mathrm{stKBM} < \beta_\mathrm{crit}^\mathrm{KBM} < \beta_\mathrm{crit}^\mathrm{MHD}$. 

Regarding geometric conditions needed for the emergence of stKBMs, low background magnetic shear $\hat{s}$ is essential for allowing very broad eigenmodes to exist along the field line -- a characteristic feature of the stKBM -- but by itself is insufficient to fully explain their presence. This is evidenced by the notable absence of stKBMs in the low-shear tokamak considered here (see also Ref.~\cite{Mulholland25}, where a similar low-shear tokmak study is presented using a lower pressure gradient), which shares a similar background magnetic shear as W7-X of $\hat{s} \approx 0.1$. Despite this ostensible similarity, it is important to highlight that $\hat{s}$ is an averaged quantity and thus does not provide local information of the field-line geometry; the local differences that persist between these configurations must be factored in when aiming to explain stKBM emergence. These local quantities are those pertaining to FLR stabilization, the magnetic-field-line curvature, and local shear, each of which are discussed here.

In the simplified KBM theory outlined here, $\hat{s}$ does not appear explicitly, but plays a role indirectly via magnetic-shear stabilization (geometric effect) and FLR stabilization (arising from the Bessel functions; geometric and plasma effect), both of which depend on the metric-tensor quantity $g^{yy}(\theta)$. As an instructive example, consider a circular tokamak equilibrium (with $\beta = \alpha_\mathrm{MHD} = 0$, for simplicity) where this quantity is described by $g^{yy}(\theta) = 1 + \hat{s}^2 \theta^2$, such that the amount of suppression from magnetic-shear and FLR effects increases quadratically with ballooning angle $\theta$; for high values of $\hat{s}$, this stabilizing effect quickly becomes substantial such that eigenmodes are restricted to exist near the outboard midplane $\theta \approx 0$, and no substantial amplitude is expected at higher $|\theta|$. In contrast, low values of $\hat{s}$ lead to weak suppression at $\theta \neq 0$, such that modes can have substantial amplitude farther along the field line (higher $|\theta|$) and are not restricted to the outboard midplane. 

The magnetic-field-line curvature -- manifest as $\Omega_\kappa$ -- is a prominent geometric term included in the KBM theory developed here. Bad-curvature regions ($\mathcal{K}_y < 0$) act to drive the mode, and so play an important role in determining KBM stability. A notable distinction between the curvature observed in W7-X and the tokamak is its spatial variation along the field line. For example, in a single poloidal turn, W7-X (KJM) has five significant bad-curvature wells, while in the same field-line domain, the tokamak has one. The higher spatial variation of curvature in W7-X ensures the mode (near marginality) is driven consistently along the field line by densely spaced neighboring bad-curvature wells; this feature, combined with the lack of strong FLR suppression due to low shear, likely aids the mode in remaining correlated over large portions of the field-line domain. Such non-localized modes may be more prone to early (low-$\beta$) destabilization due to their ability to have significant amplitude over substantial portions of the field-line domain without significantly bending the field line in any particular location; a lack of field-line bending from the mode implies a lack of significant stabilizing field-line tension in response. For $\beta_\mathrm{crit}^\mathrm{stKBM} < \beta < \beta_\mathrm{crit}^\mathrm{KBM}$, the stKBM eigenfunction is shown to become increasingly localized with increasing pressure, such that an increase in field-line tension should be experienced by the mode. This balance of increasing drive with increasing restorative stabilization may explain the observed trend of the stKBM's approximately constant growth rate for $\beta_\mathrm{crit}^\mathrm{stKBM} < \beta < \beta_\mathrm{crit}^\mathrm{KBM}$. Beyond this region, mode narrowing can no longer significantly increase, and the destabilizing drive largely overwhelms the field-line tension; this marks the transition to the fast-growing conventional KBM. These observations support the notion that important characteristics of the stKBM can be explained by employing a simplified non-resonant picture (e.g., the eigenfunction shape and eigenvalue trend), but it should be reiterated that a comprehensive understanding of the stability of this mode requires accounting for kinetic-resonance effects. 

The low-shear tokamak also hosts very broad KBM eigenfunctions, and yet does not produce an stKBM; instead, a fast-growing KBM is immediately present at the lowest $\beta$. This absence of an stKBM is potentially rooted in a combination of geometric characteristics, including -- but not limited to -- the aforementioned background-magnetic shear and magnetic-field-line curvature. Another geometric quantity that may factor in is the local magnetic shear along the field line, defined by $\hat{s}_\mathrm{loc} = \partial(g^{xy}/g^{xx})/\partial \theta$. Although this term is not directly present in the simplified KBM model, a related term is: $\partial g^{yy}/\partial \theta$. Once again, an instructive example to connect these quantities is the circular tokamak equilibrium, for which $g^{yy} = 1 + (g^{xy})^2$, and $g^{xx}=1$. In this case, $\hat{s}_\mathrm{loc} \propto \partial g^{yy}/\partial \theta$. This fluctuating quantity (along $\theta$) has notably higher magnitude (local-shear spikes) for a large portion of the field-line domain in W7-X compared with the low-shear tokamak, despite these configurations having comparable background magnetic shear. Such a feature in W7-X -- possibly in combination with the field-line-tension stabilization described above -- may aid in suppressing the stKBM after it has first become unstable at $\beta_\mathrm{crit}^\mathrm{stKBM}$ such that its growth rate remains approximately constant before reaching $\beta_\mathrm{crit}^\mathrm{KBM}$, and would at least in part explain the large separation between $\beta_\mathrm{crit}^\mathrm{stKBM}$ and $\beta_\mathrm{crit}^\mathrm{KBM}$ observed in W7-X. The absence of such substantial local-shear spikes in the low-shear tokamak may explain the immediate manifestation of a strongly driven conventional KBM at low $\beta$ rather than an stKBM. 

The effects of these geometric characteristics on (st)KBM behavior in stellarator geometry will be studied in greater detail in future work; specifically, the influence of local/background magnetic shear and Shafranov shift.  

\newpage

\section{Conclusions} \label{sect:conclusion}

One focus of magnetic-confinement-fusion research is to improve on the energy-confinement time which is largely limited by energy losses from turbulent transport. As fusion devices strive to achieve high performance (high pressure), regimes with subtantial $\beta$ must be studied and understood. The pressure-gradient driven instabilities relevant in such scenarios are then of particular interest to address, such that they can be controlled or mitigated. This work aims to expand the current understanding of one such instability -- the KBM -- in general toroidal geometry (i.e., in tokamaks and stellarators) that includes kinetic effects such as magnetic-drift resonances and FLR stabilization. 

The KBM theory developed here (see section~\ref{sect:simplified_kbm_theory}) relies on several simplifications: certain physics effects are neglected, such as particle trapping, parallel-magnetic-field fluctuations $\delta B_\parallel$, Landau resonances, and collisions; the absence of $\delta B_\parallel$ prompts the use of a magnetic-curvature-drift approximation of $\omega_{\nabla B} \equiv \omega_\kappa$; and only eigenmodes centered at $k_x = 0$ are considered. The resulting KBM eigenmode equation (\ref{eq:dimensionless_kbm_eqn_1}) depends nonlinearly on the KBM eigenvalue, and thus is reformulated as a nonlinear eigenvalue problem (see section~\ref{sect:kbm_eigenvalue_problem}); the KBM eigenvalues are first obtained, followed by treating the system as a special case of a linear eigenvalue problem, which readily yields the KBM eigenvectors. Despite the aforementioned simplifications, this KBM model performs well in rapidly predicting the correct eigenvalue trends and KBM eigenfunctions in a variety of scenarios (W7-X stellarator; high- and low-shear circular tokamak), as compared with \textsc{Gene} simulations (see section~\ref{sect:results}).

The numerical implementation of this theory has resulted in the KBM eigenvalue yielder (\textsc{Key}) code, capable of calculating self-consistent eigenvalues and eigenvectors for (st)KBMs along the field line in any flux-tube geometry, where the only required inputs are flux-tube-geometry and plasma parameters. Applying the \textsc{Key} code to both stellarator and tokamak geometry elucidates the required conditions for the manifestation of stKBMs (discussed in sect:~\ref{sect:results_discussion}): wave-particle resonances involving passing ions subject to the magnetic-curvature drift must be accounted for, which can heavily reduce the destabilization threshold of the KBM relative to the iMHD threshold for $\eta_\mathrm{i} \neq 0$; low background magnetic shear $\hat{s}$ is necessary to allow for the existence of broad eigenmodes that remain correlated for large extents along the field line; other factors that may play an important role in determining stKBM emergence are the magnetic-field-line curvature and the local magnetic shear along the field-line domain (see discussion in section~\ref{sect:results_discussion}). A closer examination of these geometric influences on (st)KBM behavior is reserved for future work. 

Despite its simplicity, this model shows great promise in a variety of aspects: it can be used as a predictive tool for determining stKBM manifestation in general toroidal geometry, which may aid in the development of reduced models for predicting turbulent transport at high $\beta$; the simplified theory in \textsc{Key} provides a basic framework to help elucidate the essential physics of (st)KBMs, enhancing the current understanding of these instabilities and their relationship with geometry; finally, the speed, affordability, and fidelity of \textsc{Key} allow for its utilization in turbulence-optimization routines for both shaped-tokamaks and stellarators aiming to achieve and sustain high-$\beta$ plasma. Thus, the results of this work may help inform the design of high-$\beta$-turbulence-optimized fusion reactors. 

\newpage

\begin{acknowledgments}

\textit{Acknowledgments} \\

The authors are grateful to P. Costello, L. Podavini, C. C. Hegna, L. J. F. Hol, F. Wilms and C. D. Stephens for valuable discussions.\\

This work has been carried out within the framework of the EUROfusion Consortium, funded by the European Union via the Euratom Research and Training Programme (Grant Agreement No 101052200 — EUROfusion). Views and opinions expressed are however those of the author(s) only and do not necessarily reflect those of the European Union or the European Commission. Neither the European Union nor the European Commission can be held responsible for them. \\

This work was supported by the Office of Fusion Energy Sciences, U.S. Department of Energy, under Grant No. DEFG02-89ER53291.

\end{acknowledgments}

\bibliographystyle{apsrev4-1}
\bibliography{refs}

\end{document}